\documentclass[11pt,a4paper]{article}
\pdfoutput=1
\usepackage{jheppub}
\usepackage{graphicx,amsmath,amssymb,amscd}
\usepackage{latexsym}
\usepackage{bm}
\usepackage{booktabs}
\usepackage{mathrsfs}
\usepackage{float}
\usepackage{fancybox}
\usepackage{braket}
\usepackage{ascmac}
\usepackage{fancybox,pifont,layout}
\usepackage{amsmath,amsthm}
\usepackage{color}
\allowdisplaybreaks[3] 
\title{\boldmath Realization of lepton masses and mixing angles from point~interactions in an extra dimension}

\author[a]{Yukihiro Fujimoto,}
\author[b,c,1]{Kenji Nishiwaki,\note{ Corresponding author.}}
\author[d]{Makoto Sakamoto,}
\author[e]{and Ryo Takahashi\,}

\affiliation[a]{Department of Physics, Osaka University,\\
			Machikaneyama-Cho 1-1, Toyonaka 560-0043, Japan}
\affiliation[b]{Regional Centre for Accelerator-based Particle Physics,
			Harish-Chandra Research Institute,\\ Chhatnag Road, Jhusi, Allahabad 211 019, India}
\affiliation[c]{School of Physics, Korea Institute for Advanced Study,\\
			85 Hoegiro, Dongdaemun-gu, Seoul 130-722, Republic of Korea}
\affiliation[d]{Department of Physics, Kobe University,\\
			1-1 Rokkodai, Nada, Kobe 657-8501, Japan}
\affiliation[e]{Graduate School of Science and Engineering, Shimane University,\\
			1060 Nishikawatsu, Matsue, Shimane 690-8504, Japan}

\emailAdd{fujimoto@het.phys.sci.osaka-u.ac.jp}
\emailAdd{nishiken@kias.re.kr}
\emailAdd{dragon@kobe-u.ac.jp}
\emailAdd{takahashi@riko.shimane-u.ac.jp}

\abstract{
We investigate a model on an extra dimension $S^1$ where plenty of effective boundary points described by point interactions (zero-thickness branes) are arranged.
After suitably selecting the conditions on these points for each type of five-dimensional fields, we realize the tiny active neutrino masses, the charged lepton mass hierarchy, and lepton mixings with a CP-violating phase, simultaneously.
Not only the quark's but also the lepton's configurations are generated in a unified way with acceptable accuracy, with neither the see-saw mechanism nor symmetries in Yukawa couplings, by suitably setting the model parameters, even though their flavor structures are dissimilar each other.
One remarkable point is that a complex vacuum expectation value of the five-dimensional Higgs doublet in this model becomes the common origin of the CP violation in both quark and lepton sectors.
The model can be consistent with the results of the precision electroweak measurements and Large Hadron Collider experiments.
}

\begin{document} 
\maketitle
\flushbottom


\section{{\large Introduction}}

While the gauge sector of the Standard Model (SM) had been completed by the discovery of the Higgs boson~\cite{:2012gk,:2012gu}, the actual structure of the Yukawa sector generating fermion mass terms is still concealed.
The SM Yukawa terms {describe} the nature very well, but we should introduce three similar copies for realization of the three generations and a large number of parameters describing the masses and the mixing-related issues of {both the quarks and the leptons. The values of the parameters} are simply input parameters and are never determined by dynamics.
To make matters worse, even after we accept the above points, the naturalness issue frightens us.
Only within the quarks, $10^5$-order hierarchy in the input Yukawa couplings should be generated by hand.
When we try to include the lepton sector, situations get to be deteriorated as the magnitude of fine-tuning should be enlarged to at least $10^{11}$ orders since the observed active neutrinos has sub-eV tiny masses.\footnote{
Note that a recent upper bound on the sum of neutrino masses by Planck experiment is $0.23\,\text{eV}$~\cite{Ade:2013zuv}, and the possibility that the lightest active neutrino is a massless particle is not discarded yet.
}
Apart from the naturalness issue, another mystery is there with respect to the mixing structure of the fermions.
The {Cabbibo-Kobayashi-Maskawa} (CKM) matrix~\cite{Kobayashi:1973fv} showing the quark mixing phenomena consists of three small mixing angles with a CP-violating phase.
On the other hand, two of the lepton mixing angles{, $\theta_{12}$ and $\theta_{23}$,} in the {Pontecorvo-Maki-Nakagawa-Sakata} (PMNS) matrix~\cite{Maki:1962mu,Pontecorvo:1967fh} are large, while the remaining one $\theta_{13}$ is small but has a non-zero value, declared by recent experiments~\cite{Abe:2011sj,Adamson:2011qu,Abe:2011fz,An:2012eh,Ahn:2012nd} and following global analyses~\cite{Tortola:2012te,Fogli:2012ua,GonzalezGarcia:2012sz,Capozzi:2013csa}.
Note that the existence of CP-violating phases in the lepton sector is unknown to date.

One expects that these characteristics would be naturally explained by candidates describing the physics beyond the SM, e.g., {through} the see-saw mechanism for a fascinating explanation for the minuscule neutrino masses~\cite{Minkowski:1977sc,Yanagida:see-saw,Gell-Mann:see-saw,Mohapatra:1979ia,Schechter:1980gr,Schechter:1981cv}.
Here, related to the Yukawa structure, we mention two important issues.
One is that the number of matter generation is a kind of {``topological number''.}
Within four-dimensional (4d) spacetime, we mainly rely on the gauge anomaly cancellation for discussing this point, e.g., in the following extended electroweak (EW) gauge theories,
$SU(3)_W \times U(1)_Y$~\cite{Pisano:1991ee,Frampton:1992wt} or $SU(2)_W \times U(1)_Y \times U(1)_{B-L}$~\cite{Babu:1989tq,Babu:1989ex}.
{The other point is the Higgs mechanism.}
{Electroweak symmetry breaking (EWSB)} via the Higgs mechanism leads to {the} generation of the current mass of fermions through Yukawa couplings.
But the realization of EWSB within the SM would be somewhat {\it ad hoc} since it is just assumed {and has} no relation to other dynamics directly.

When we pursue a suitable answer for problems in flavor, one of the engaging ways is considering a hidden structure of extra dimensions~\cite{ArkaniHamed:1998vp,ArkaniHamed:1999dc,Dvali:1999cn,Yoshioka:1999ds,Mohapatra:1999zd,Grossman:1999ra,Dvali:2000ha,Gherghetta:2000qt,Huber:2000ie,Libanov:2000uf,Frere:2000dc,Neronov:2001qv,Frere:2001ug,Kaplan:2001ga,Cremades:2004wa,Parameswaran:2006db,Gogberashvili:2007gg,Kaplan:2011vz,Fujimoto:2013xha,Abe:2013bca}{. A large} variety of solitonic objects spreading {around} the extra directions explains the number of generations through their topological numbers without violating 4d Poincar\'e symmetry and {the} localization of matter profiles lead us to a natural explanation of mass hierarchies.
In this paper, we focus on the direction proposed in Refs.~\cite{Fujimoto:2012wv,Fujimoto:2013ki},
where we think of generalized boundary conditions (BC's) described by point interactions (zero-thickness {branes}) in the bulk space $S^1$~\cite{Hatanaka:1999ac,Nagasawa:2002un,Nagasawa:2003tw,Nagasawa:2005kv}.
By selecting the BC's for five-dimensional (5d) fields suitably, we naturally realize three-degenerated localized fermion profiles, EWSB and the SM gauge boson configuration in the zero-mode sector simultaneously without suffering from any EW precision data and the recent Higgs results at the Large Hadron Collider.
As discussed in Ref.~\cite{Fujimoto:2013ki}, by choosing parameters in the model appropriately, the whole configurations of the quark sector including the number of the quarks, the quark masses and the mixing angles with one CP phase are explained with good precision.
Here, we note that one additional EW singlet scalar is introduced with an {extra-dimension coordinate-dependent }exponential vacuum expectation value (VEV)~\cite{Sakamoto:1999yk,Ohnishi:2000hs,Hatanaka:2000zq,Sakamoto:1999ym,Sakamoto:1999iv,Matsumoto:2001fp,Sakamoto:2001gn,Coradeschi:2007gb,Burgess:2008ka,Haba:2009uu} in the Yukawa sector to enhance the hierarchy in the elements of the Yukawa matrices.
Interestingly in the model, a complex degrees of freedom for the CP-violating phase in the CKM matrix is supplied via a twisted boundary condition on the Higgs doublet, whose VEV is also determined as a minimization of the corresponding effective potential, as the realization of a complex VEV.
After calculating overlap integrals between the position-dependent VEV's and {localized quark profiles}, desirable forms in the Yukawa matrices are produced effectively.

{Our main purpose of this paper is to answer a natural question whether or not the above idea is applicable to the lepton sector. 
As we mentioned before, we should explain the sub-eV neutrino masses with the two large mixing angles, as well as the quark masses and mixing angles, in a unified manner.}
Situations are apparently different from those of the quark sector and we need to find a suitable configuration of the system.
We have found that all the lepton properties can be described with acceptable precision after carefully fitting effective parameters.
One of the captivating points of the model is predicting the CP violation effect in the lepton sector, whose origin is identified as the complex VEV of the doublet Higgs and then
we can predict the magnitude by use of the information on the quark sector.

This paper is organized as follows.
In section~\ref{section:lepton}, first we summarize the way of our model construction with point interaction briefly and
subsequently, we apply the idea to the lepton sector and discuss the validity from quantitative and qualitative points of views.
Section~\ref{section:summary} is devoted to summary and further discussions.
{In Appendix A, we review the quark sector analysis. In Appendix B, we discuss how the form of the mass matrix is restricted by the geometry of the extra dimension. }


\section{{\large Lepton flavor structure from point interactions} \label{section:lepton}}

{In this section, we briefly overview our concept and idea for generating the three-generation structure, the hierarchy and mixing in the lepton mass matrices, switching on {EWSB} and CP violation effect. Then, we show our {numerical} results showing how much our mechanism works correctly in a focus point of parameters.}
The model is a 5d gauge theory on a circle $S^1$ with point interactions and contains one-generation fermion for each $SU(2)_W$ eigenstate at the level of 5d action.
By imposing the suitable BC's for the fields at the positions of the point interactions, we obtain a three-generation structure in chiral massless zero modes.
The rather mass hierarchy in the charged leptons comes from the intercorrection between the localization of the chiral massless zero modes toward the boundary points and the {extra-dimension coordinate-dependent} VEV of the gauge singlet scalar.
On the other hand, minuscule masses {of} the neutrinos are realizable by extremely localized profiles of the neutrinos.
A significant feature appears in the lepton flavor mixing since it is determined by the configuration of the point interactions in our model.
The origin of the physical CP phase cannot be the Yukawa couplings themselves since our model consists of {one-generation} fermions {and hence field redefinitions can make them real in general}.
It comes from the oscillatory VEV of the Higgs doublet, which includes a complex degree of freedom.


{\subsection{{Action and configuration of point interactions}}
\label{subsection-model}}

\hspace{1.5em}
Let us introduce the action and the geometry of the extra dimension. 
{We mainly focus on the lepton sector. The analysis of the quark sector will briefly be summarized in Appendix A.}
We consider a 5d {$SU(3)_C \times SU(2)_W \times U(1)_Y$} gauge theory on $S^1$ with point interactions. The action consists of one-generation 5d lepton fields with the Higgs doublet, the gauge singlet scalar and the Yukawa sector. 
	\begin{align}
	&S=S_{{\rm lepton}}+S_{{\rm Higgs}}+S_{{\rm singlet}}+S_{{\rm Yukawa}}^{({\rm lepton})},\\[4pt]
	&S_{{\rm lepton}}=\int d^4 x \int^{L}_{0}dy \left[\overline{L}(x,y)\Bigl(i\Gamma^{N}D_{N}^{(L)}+M_{L}\Bigr)L(x,y)\right.\nonumber\\
	&\left.\hspace{3em}+\overline{{\cal N}}(x,y)\Bigl( i\Gamma^{N} {\partial_{N}}+M_{{\cal N}}\Bigr) {\cal N}(x,y) +\overline{E}(x,y)\Bigl( i\Gamma^{N}D_{N}^{(E)}+M_{ E}\Bigr) E(x,y)\right],\label{Slepton}\\
	&S_{{\rm Higgs}}=\int d^4 x \int^{L}_{0}dy \left[H^{\dagger}(x,y)\Bigl(D^{N}D_{N}+M^2\Bigr)H(x,y)-\frac{\lambda}{2}\Bigl(H^{\dagger}(x,y)H(x,y)\Bigr)^2\right],\\[4pt]
	&S_{{\rm singlet}}=\int d^4 x \int^{L}_{0}dy \left[\Phi^{\dagger}(x,y)\Bigl(\partial^{N}\partial_{N}-M_{\Phi}^2\Bigr)\Phi(x,y)-\frac{\lambda_{\Phi}}{2}\Bigl(\Phi^{\dagger}(x,y)\Phi(x,y)\Bigr)^2\right],\label{5daction_singlet} \\
	&S_{{\rm Yukawa}}^{({\rm lepton})}=\int d^4 x\int^{L}_{0}dy \left[ \Phi\Bigl(-{\cal Y}^{({\cal N})}\overline{ L}(i\sigma_{2}H^{\ast}){\cal N}-{\cal Y}^{(E)}\overline{L}HE\Bigr)+({\rm h.c.})\right]{,}\label{leptonYukawa}
	\end{align}
where we denote $L(x,y)$ as an $SU(2)_{W}$ doublet lepton, ${\cal N}(x,y)$ as an $SU(2)_{W}$ singlet neutrino, $E(x,y)$ as an $SU(2)_{W}$ singlet charged lepton,  $H(x,y)$ as the Higgs doublet and $\Phi(x,y)$ as an gauge singlet scalar field, respectively.
{$D_{N}^{(\Psi)}\,(\Psi=L,E)$ shows the corresponding 5d covariant derivatives and note that} {${\cal N}$ does not couple to the gauge fields since it corresponds to the right-handed neutrino in 4d point of view}.
The variable $x^{\mu}$ $(\mu=0,1,2,3)$ indicates the coordinate of the 4d Minkowski spacetime and $y$ is the coordinate of the extra dimension {with the circumference of $S^1$, $L$}.
{$N$ (and $M$) runs among $\mu,y$ as a 5d Lorentz index.}
The 5d metric {is} chosen as $\eta_{MN}={\rm diag}(-1,+1,+1,+1,+1)$ and the Clifford algebra as $\{\Gamma_{M},\Gamma_{N}\}=-2\eta_{MN}$ with adopting the representation of the gamma matrices as $\Gamma_{\mu}=\gamma_{\mu}$ and $\Gamma_{y}=-i\gamma_{5}=\gamma^{0}\gamma^{1}\gamma^{2}\gamma^{3}$.
{We note that our theory is defined in the segment $[0,L]$.}

$M_{\Psi}\ (\Psi=L, {\cal N}, E)$ {in Eq.~(\ref{Slepton})} denotes {the} bulk mass for the fermion and we put the sign as
	\begin{align}
	&M_{L}<0,\label{signbulkmassL}\\
	&M_{{\cal N}}>0,\label{signbulkmassN}\\ 
	&M_{E}<0,\label{signbulkmassE}
	\end{align}
for {our purpose. We explain the reason of this adoption later in this section.} 
Here, we omit the action for the {$SU(3)_C \times SU(2)_W \times U(1)_Y$} gauge fields {since their situations become a simple Universal-Extra-Dimension-like profile still in our strategy}.

Note that {the discrete symmetry, $H\rightarrow -H$, $\Phi\rightarrow -\Phi$, is} introduced to forbid the terms $\overline{L}(i\sigma_{2}H^{\ast}){\cal N}$, $\overline{L}HE$, $\Phi \overline{L}L$, $\Phi \overline{{\cal N}}{\cal N}$, $\Phi \overline{E}E$ {for simplicity}. We also simply ignore the term $(H^{\dagger}H)(\Phi^{\dagger}\Phi)$ in this model.
	\footnote{
The term $(H^{\dagger}H)(\Phi^{\dagger}\Phi)$ can be a source of gauge universality violation in this model and its effect was investigated in Ref \cite{Fujimoto:2012wv}. See, its {Appendices} A and B.}
We also mention that although the Yukawa couplings {${\cal Y}^{(\mathcal{N})}$} and ${\cal Y}^{(E)}$ can be complex, they cannot be an origin of the CP phase of the PMNS matrix since the model consists of only {one-generation} leptons in 5d point of view. The number of the 5d Yukawa couplings is not enough to produce a CP phase in the PMNS matrix. A source of the CP phase is the VEV of the Higgs in the model and will be discussed in subsection~\ref{subsection-CPphase}.

\begin{figure}[t]
	\begin{center}
	\includegraphics[width=0.9\columnwidth]{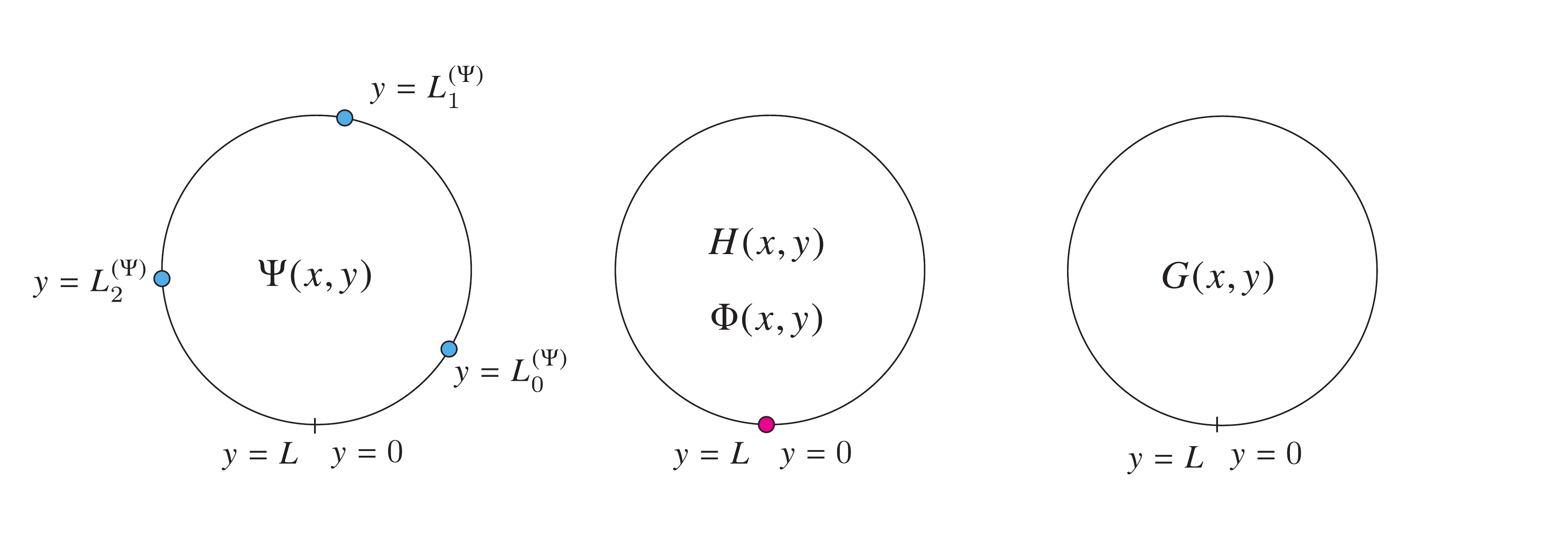}
	\end{center}
	\vspace{-0.5cm}\caption{{\small A schematic figure of the configuration of the point interactions. The fermions feel three point interactions, the gauge singlet scalar and the Higgs doublet feel one, the gauge fields do not feel the point interactions in the model. }}
	\label{figure-configuration}
	\end{figure}

In the case of {an} extra dimension scenario, not only the action but also the geometry of the extra dimension is important.
In the model, the geometry consists of $S^1$ with point interactions.
Every fermion field ($L,\, {\cal N},\, E$) feels three point interactions at the positions $y=L_{i}^{(\Psi)}$ ($\Psi=L, {\cal N}, E; \ i=0,1,2,3$), respectively.
On the other hand, the Higgs {doublet $H$} and the gauge singlet scalar {$\Phi$} feel one point interaction at $y=0$.
{Gauge fields do not feel any point interaction, where the usual periodic boundary condition is chosen.}
A schematic figure is depicted in {Fig.}~\ref{figure-configuration}.

\begin{figure}[t]
	\begin{center}
	\includegraphics[width=1.1\columnwidth]{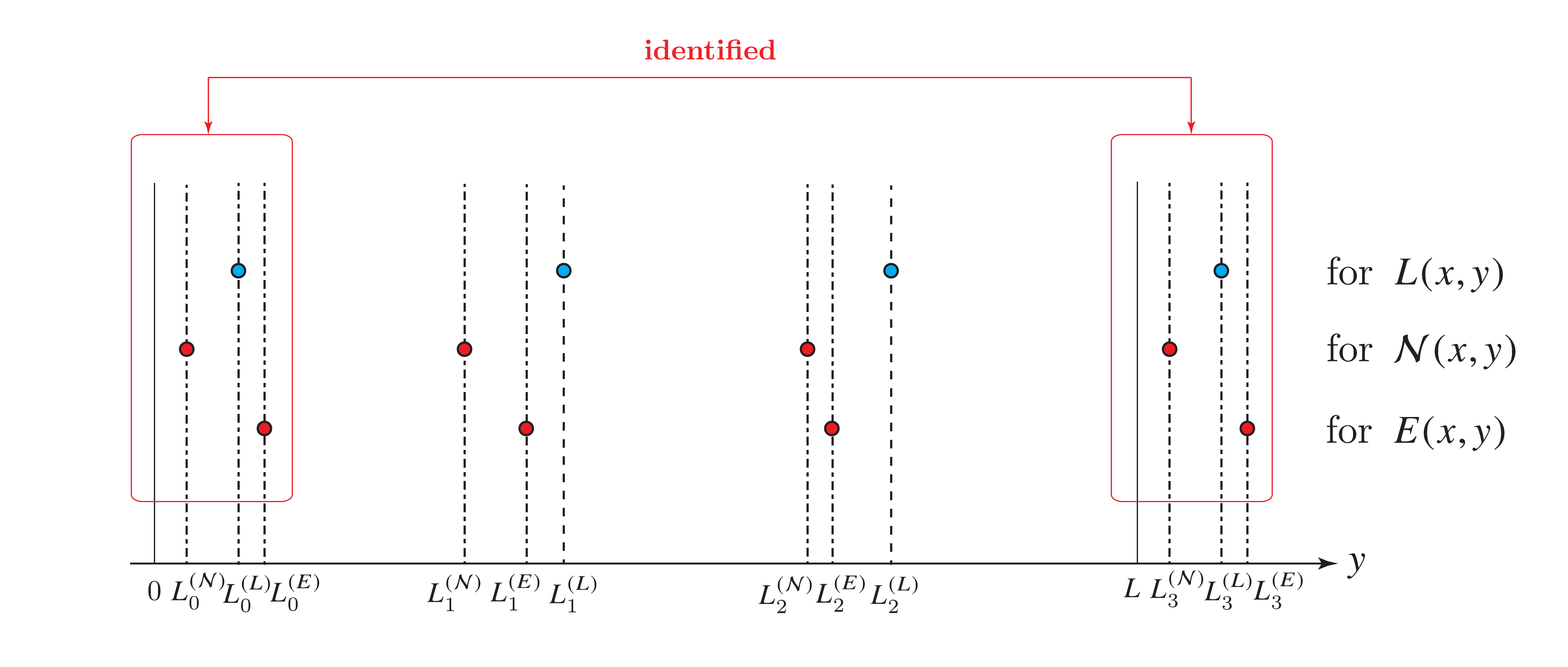}
	\end{center}
	\vspace{-0.5cm}\caption{{\small A schematic figure of the configuration of the point interactions. }}
	\label{figure-positions}
	\end{figure}

Because of the point interactions, every field feels boundary conditions at {its own} positions. We impose the characteristic BC's for the fermions and the singlet scalar {at} which the flow of the probability current {is} not allowed through the point interactions. It was pointed out in Ref.~\cite{Fujimoto:2012wv} that the general BC's in the situation is given by $\Psi_{R}=0$ or $\Psi_{L}=0$ at $y=L_{i}^{(\Psi)}$ for the fermion where the indices $R$ and $L$ denote the 4d chirality as $\Psi_{R}\equiv \left(\frac{1+\gamma_{5}}{2}\right)\Psi,\ \Psi_{L}\equiv \left(\frac{1-\gamma_{5}}{2}\right)\Psi$. Then we impose the following BC's.
	\begin{align}
	L_{R}&=0\hspace{3em}{\rm at}\ \ \  y=L_{0}^{(L)},L_{1}^{(L)},L_{2}^{(L)},L_{3}^{(L)}\ , \label{LBC}\\
	{\cal N}_{L}&=0\hspace{3em}{\rm at}\ \ \ y=L_{0}^{({\cal N})},L_{1}^{({\cal N})},L_{2}^{({\cal N})},L_{3}^{({\cal N})}\ ,\label{NBC}\\
	E_{L}&=0\hspace{3em}{\rm at}\ \ \ y=L_{0}^{(E)},L_{1}^{(E)},L_{2}^{(E)},L_{3}^{(E)}\ .\label{EBC}
	\end{align}
{The BC's for the partners of the different 4d chirality just come from the 5d Dirac equation.} In this model, we choose the following {order of the point interactions of the fermions:}
	\begin{align}
	&0<L_{0}^{({\cal N})}<L_{0}^{(L)}<L_{1}^{({\cal N})}<L_{1}^{(L)}<L_{2}^{({\cal N})}<L_{2}^{(L)}<L_{3}^{{({\cal N})}}<L_{3}^{(L)}, \label{nuconfiguration}\\
	&0<L_{0}^{(L)}<L_{0}^{(E)}<L_{1}^{(E)}<L_{1}^{(L)}<L_{2}^{(E)}<L_{2}^{(L)}<L_{3}^{(L)}<L_{3}^{(E)}.\label{econfiguration}
	\end{align}
We also explain the reason of this adoption later in this section.
We should mention that the following positions should be identified since the geometry of the extra dimension is $S^1$ with the period $L$.
	\begin{align}
	L &\sim 0,\\
	L_{3}^{({\cal N})} &\sim L_{0}^{({\cal N})},\\
	L_{3}^{(L)} &\sim L_{0}^{(L)},\\
	L_{3}^{(E)} &\sim L_{0}^{(E)},
	\end{align}
the situation of which is depicted in Fig.~\ref{figure-positions}.

For the singlet scalar, the general BC is given by the Robin BC~\cite{Fujimoto:2011kf}:
	\begin{align}
	\left\{
	\begin{array}{l}
	\Phi(x,0)+L_{+}\partial_{y}\Phi(x,0)=0,\\
	\Phi(x,L)-L_{-}\partial_{y}\Phi(x,L)=0,
	\end{array}
	\right.
	\hspace{3em}(-\infty\leq L_{\pm}\leq +\infty),\label{PhiBC}
	\end{align}
where $L_{\pm}$ {are parameters which describe }the BC.
Note that by suitably choosing the parameters $L_{\pm}$ and other parameters in Eq.~(\ref{5daction_singlet}), the form of the VEV gets to be hierarchical along the $y$ direction.
Such a situation is preferable for generating the large hierarchy in the lepton mass matrices.

{For} the Higgs doublet and the gauge fields, the flow of the probability current is allowed {on the circle $S^1$}. We impose the BC's as
	\begin{align}
	&\left\{
	{\begin{array}{l}
	H(x,L) = e^{i\theta}H(x,0),\\
	\partial_y H(x,L) = e^{i\theta} \partial_y H(x,0),
	\end{array}} \hspace{5em}(-\pi<\theta\leq \pi),
	\right. \label{HBC} \\[0.3cm]
	&{\left\{
	\begin{array}{l}
	\mathcal{G}_{N}(x,L)={\cal G}_{N}(x,0),\\
	\partial_{y} {{\cal G}_{N}} (x,L) = \partial_{y}{\cal G}_{N}(x,0),
	\end{array}  \hspace{3em}({\cal G}_{N}= G_{N},\ W_{N},\ B_{N}),
	\right.}\label{GBC}
	\end{align}
where $\theta$ is a phase parameter which specifies the twisted BC, whose complex degree of freedom is just the origin of CP violation in our model. 
{$G_{N}$, $W_{N}$ and $B_{N}$ indicate a $SU(3)$ gauge field, a $SU(2)$ gauge field and a $U(1)$ gauge field, respectively.}

We should emphasize that {all the BC's} are consistent with the 5d gauge invariance. {In particular,} the BC's~{(\ref{LBC})--(\ref{EBC})} do not break the 5d gauge symmetry since the BC's for the fermion are given by the Dirichlet BC, which is manifestly invariant under the 5d gauge transformation.\footnote{
	The higher-dimensional gauge invariance of the system is important when we discuss the unitarity in the scattering processes of KK particles~\cite{SekharChivukula:2001hz,Abe:2003vg,Chivukula:2003kq,Csaki:2003dt,Ohl:2003dp,Abe:2004wv,Sakai:2006qi,Nishiwaki:2010te}.
	}

\subsection{Three generations}\label{subsection-generations}

\begin{figure}[t]
	\begin{minipage}{18cm}
	\includegraphics[width=0.99\columnwidth]{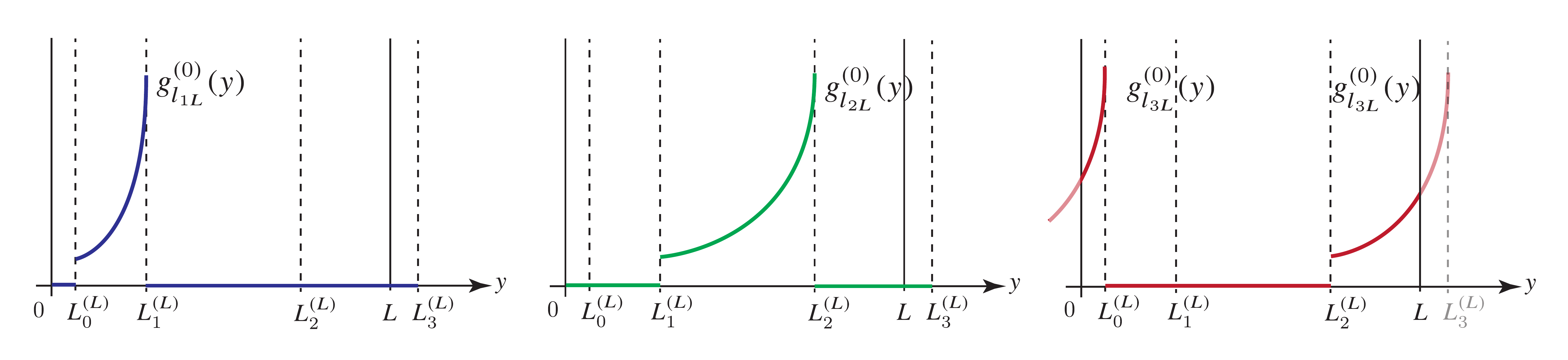}
	\par
		\vspace{0.0cm}
		{\footnotesize{(a) The profiles of triply-degenerated zero modes $g_{l_{iL}}^{(0)}$ with $M_{L}>0$. (The profiles of $f_{\nu_{iR}}^{(0)}$, $f_{e_{iR}}^{(0)}$ with $M_{\cal N}$, $M_{E}<0$.)}}
	\end{minipage}\\
	\begin{minipage}{18cm}
	\includegraphics[width=0.99\columnwidth]{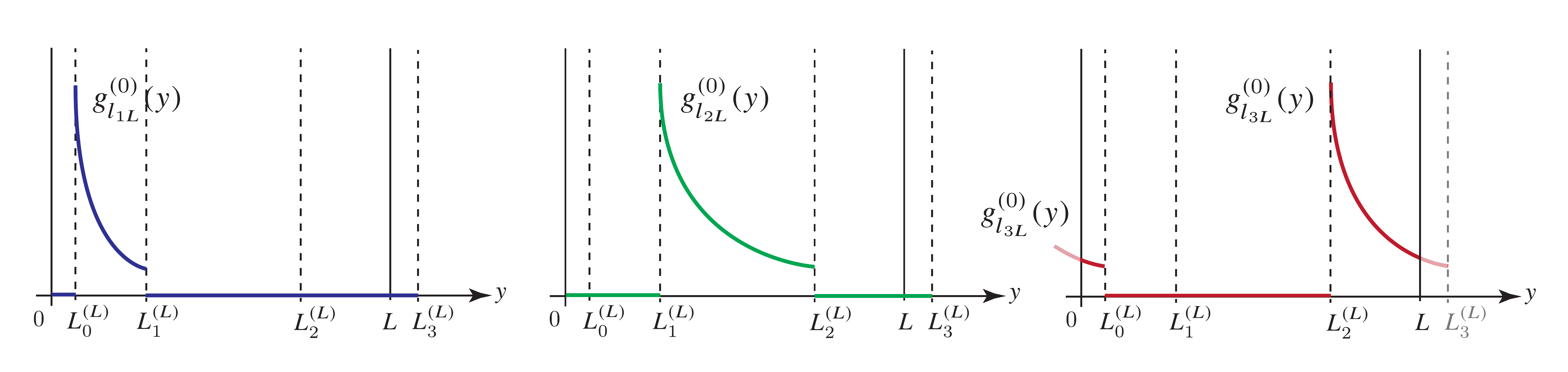}
	\par
		\vspace{0.0cm}
		{\footnotesize{(b) The profiles of triply-degenerated zero modes $g_{l_{iL}}^{(0)}$ with $M_{L}<0$. (The profiles of $f_{\nu_{iR}}^{(0)}$, $f_{e_{iR}}^{(0)}$ with $M_{\cal N}$, $M_{E}>0$.)}}
	\end{minipage}
	\caption{A schematic figure of the zero modes.}
	\label{figure_zeromodes}
	\end{figure}

In this subsection, we {explain the mechanism to} produce three generations {in} chiral massless zero modes. To see this, let us execute the Kaluza-Klein (KK) expansion for the fermions,
	\begin{align}
	\Psi(x,y)=\sum_{n}\left( \psi^{(n)}_{R}(x)f^{(n)}_{\psi_{R}}(y)+\psi^{(n)}_{L}(x)g_{\psi_{L}}^{(n)}\right), \label{KKexpansion}
	\end{align}
where $\{ f_{\psi_{R}}^{(n)}\}$ $\Bigl(\{ g_{\psi_{L}}^{(n)}\}\Bigr)$ is the eigenfunction of the hermitian operator $\mathscr{D}^{\dagger}\mathscr{D}$ $\left(\mathscr{D}\mathscr{D}^{\dagger}\right)$ and forms the complete set,
	\begin{align}
	\left\{
	\begin{array}{l}
	\mathscr{D}^{\dagger}\mathscr{D} f_{\psi_{R}}^{(n)}=m^{2}_{\psi^{(n)}}f_{\psi_{R}}^{(n)},\\[0.2cm]
	\mathscr{D}\mathscr{D}^{\dagger} g_{\psi_{L}}^{(n)}=m^{2}_{\psi^{(n)}}g_{\psi_{L}}^{(n)}{,}
	\end{array}
	\right. \hspace{3em}(\mathscr{D}\equiv \partial_{y}+M_{\Psi}, \ \mathscr{D}^{\dagger}\equiv -\partial_{y}+M_{\Psi}).
	\end{align}
The degeneracy of the eigenvalues $m_{\psi^{(n)}}^2$ {for $f^{(n)}_{\psi_{R}}$ and $g^{(n)}_{\psi_{L}}$} is ensured by quantum mechanical supersymmetry (QMSUSY)~\cite{Lim:2005rc,Lim:2007fy,Lim:2008hi,Nagasawa:2008an} as
	\begin{align}
	\left\{
	\begin{array}{l}
	\mathscr{D} f_{\psi_{R}}^{(n)}=m_{\psi^{(n)}}g_{\psi_{L}}^{(n)},\\[0.2cm]
	\mathscr{D}^{\dagger} g_{\psi_{L}}^{(n)}=m_{\psi^{(n)}}f_{\psi_{R}}^{(n)}.
	\end{array}
	\right.
	\end{align}
{Zero mode solutions} of the above {should satisfy} the following equations {since $m_{\psi^{(0)}}=0$.}
	\begin{align}
	\left\{
	\begin{array}{l}
	\mathscr{D} f_{\psi_{R}}^{(0)}=0,\\[0.2cm]
	\mathscr{D}^{\dagger} g_{\psi_{L}}^{(0)}=0.
	\end{array}
	\right.\label{zeromodeequations}
	\end{align}
{From Eq.~(\ref{zeromodeequations}), we obtain the forms of the non-trivial zero mode solutions (trivial solutions are $f^{(n)}_{\psi_{R}}=0=g^{(n)}_{\psi_{L}}$) as}
	\begin{align}
	\left\{
	\begin{array}{l}
	f_{\psi_{R}}^{(0)}\propto e^{-M_{\Psi}y},\\[0.2cm]
	g_{\psi_{L}}^{(0)}\propto e^{+M_{\Psi}y}.
	\end{array}
	\right.
	\end{align}
{Furthermore, we have to consider the BC's in Eqs.~(\ref{LBC})--(\ref{EBC}). Since the Dirichlet boundary condition does not allow any flow of the probability currents through the point interactions, the profiles can split at them. It turns out that we can obtain three generations of chiral zero modes in our configuration as follows.}
	\begin{align}
	L(x,y)&=\sum_{i=1}^{3}l^{(0)}_{i L}(x)g_{l_{iL}}^{(0)}(y)+({\rm KK\ modes}),\\
	{\cal N}(x,y)&=\sum_{i=1}^{3}\nu^{(0)}_{i R}(x)f_{\nu_{iR}}^{(0)}(y)+({\rm KK\ modes}),\\
	E(x,y)&=\sum_{i=1}^{3}e^{(0)}_{i R}(x)f_{e_{iR}}^{(0)}(y)+({\rm KK\ modes}).
	\end{align}
The degenerated zero mode solutions are depicted in {Fig.~\ref{figure_zeromodes}}. The mode functions are localized {toward} the boundary points because of the bulk mass $M_{\Psi}$ ($\Psi=L, {\cal N}, E$), the sign of whom determines the direction of {the} localization. We should note that the bulk mass controls all the related zero modes so that we cannot change the form of the degenerated zero modes, independently.

\subsection{The charged lepton mass hierarchy and tiny neutrino masses}
\label{subsection-masses}

	\begin{figure}[t]
	\hspace{-1.5cm}\includegraphics[width=1.1\columnwidth]{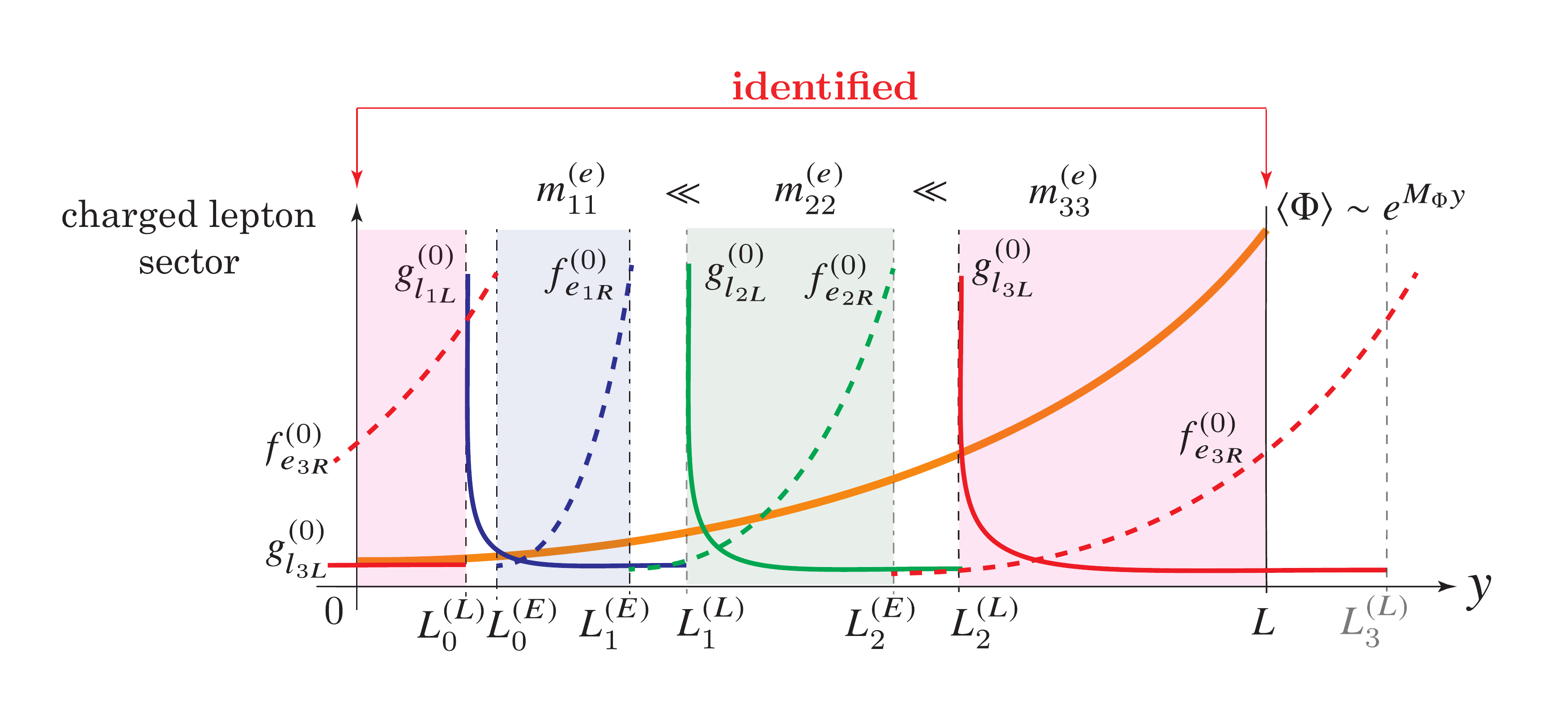}
	\vspace{-0.9cm}\caption{{\small A schematic figure of the configuration of the charged leptons. {The overlap integrals of the blue, green and pink colored region indicate the diagonal components $m_{11}^{(e)}$, $m_{22}^{(e)}$ and $m_{33}^{(e)}$ of the charged lepton mass matrix, respectively.} The exponential VEV of the gauge singlet {$\langle \Phi(y) \rangle$} produces the charged lepton mass hierarchy. }}
	\label{figure-masshierarchy_chargedlepton}
	\end{figure}

	\begin{figure}[t]
	\hspace{-1.5cm}\includegraphics[width=1.1\columnwidth]{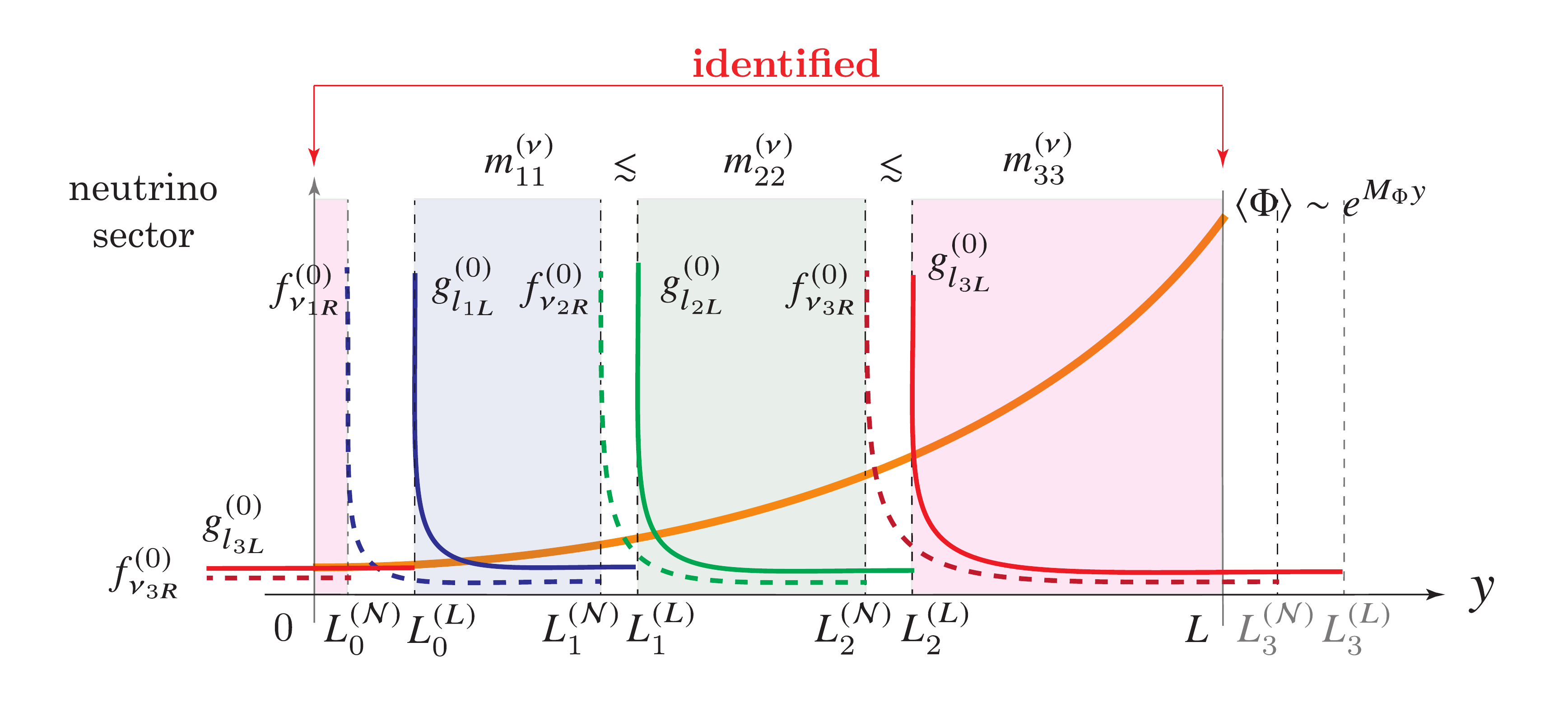}
	\vspace{-0.9cm}\caption{{\small A schematic figure of the configuration of the {neutrinos}.   {The overlap integrals of the blue, green and pink colored region indicate the diagonal components $m_{11}^{(\nu)}$, $m_{22}^{(\nu)}$ and $m_{33}^{(\nu)}$ of the neutrino mass matrix, respectively.} { Because of the immoderate localization, the overlap integrals suppressed so that degenerated tiny neutrino masses appear.}}}
	\label{figure-masshierarchy}
	\end{figure}

In this subsection, we search for the configuration of point interactions where the charged lepton mass hierarchy and tiny neutrino masses are derived. We should mention that only the Dirac mass terms are introduced to the both {charged} leptons and neutrinos in our model, which means that tiny neutrino masses appear from the geometry of the extra dimension not from the seesaw mechanism.
Under the Robin BC {in Eq.}~(\ref{PhiBC}), the gauge singlet scalar can obtain {a non-vanishing VEV $\langle\Phi(y)\rangle$} nevertheless $M_{\Phi}^2>0$~\cite{Fujimoto:2011kf}.
Moreover, the VEV {$\langle \Phi(y)\rangle$} inevitably possesses the $y$-dependence~\cite{Fujimoto:2011kf}.
	\begin{align}
	{\langle\Phi(y)\rangle=\phi(y).}
	\end{align}
After solving the minimization condition of the potential $V_{{\rm 4d}}=\int^{L}_{0}dy \left[\Phi^{\dagger}(-\partial_{y}^2+M_{\Phi}^2)\Phi+\frac{\lambda_{\Phi}}{2}(\Phi^{\dagger}\Phi)^2\right]$ with the Robin BC~(\ref{PhiBC}), {it was found that the following {form} of the VEV {is the vacuum configuration}:} 
	\begin{align}
	{\phi(y)}=\frac{\frac{M_{\Phi}}{\sqrt{\lambda_{\Phi}}}\Bigl(\sqrt{1+X}-1\Bigr)^{\frac{1}{2}}}{{\rm cn}\left(M_{\Phi}(1+X)^{\frac{1}{4}}(y-y_{0}), \sqrt{\frac{1}{2}\Bigl(1+\frac{1}{\sqrt{1+X}}\Bigr)}\right)},\hspace{3em}\biggl( X\equiv \frac{4\lambda_{\Phi}|Q|}{M_{\Phi}^4}\biggr),
	\end{align}
where {cn($y,a$) is the Jacobi's elliptic function}$, y_{0}$ and $Q$ are constants which are determined by the parameters $L_{\pm}$ of the BC's.\footnote{
{We note that this form is in the case of $Q<0$~\cite{Fujimoto:2012wv}.}
}
We can make it the exponential form by {choosing the suitable value of $L_{\pm}$},
	\begin{align}
	{\phi(y)} \sim e^{M_{\Phi}y}, \label{exponentialVEV}
	\end{align}
where we omit to show a {proportional} factor with mass dimension $3/2$.

The $y$-dependent VEV of the gauge singlet scalar {$\langle \Phi(y) \rangle$} can be a source of the charged lepton mass hierarchy since the mass matrices, which appear from the Yukawa sector {in Eq.}~(\ref{leptonYukawa}), {contain} the following overlap integrals,
	\begin{align}
	&S_{{\rm Yukawa}}^{({\rm lepton})}=\int d^4 x\int^{L}_{0}dy \left[ \Phi\Bigl(-{\cal Y}^{({\cal N})}\overline{ L}(i\sigma_{2}H^{\ast}){\cal N}-{\cal Y}^{(E)}\overline{L}HE\Bigr)+({\rm h.c.})\right]\nonumber\\
	&\quad {\supset -\int d^4 x \sum_{i,j=1}^{3}\left[ m_{ij}^{(\nu)} \overline{{\nu}_{iL}^{(0)}} \nu_{jR}^{(0)} + m_{ji}^{(\nu)}{}^{\ast} \overline{{\nu}_{jR}^{(0)}} \nu_{iL}^{(0)}\right]
- \int d^4 x \sum_{i,j=1}^{3}\left[ m_{ij}^{(e)} \overline{e_{iL}^{(0)}} e_{jR}^{(0)}
+ m_{ji}^{(e)}{}^{\ast} \overline{e_{jR}^{(0)}} e_{iL}^{(0)}\right]},
	\end{align}
with
	\begin{align}
	m_{ij}^{(\nu)}&={\cal Y}^{({\cal N})}\int^{L}_{0}dy \,{\langle h(y)\rangle^{\ast}} \,\langle \Phi(y) \rangle \,g_{l_{iL}}^{(0)}(y)\,f_{\nu_{jR}}^{(0)}(y),\label{nuoverlap}\\
	m_{ij}^{(e)}&={\cal Y}^{(E)}\int^{L}_{0}dy \,{\langle h(y)\rangle} \,\langle \Phi(y)\rangle \,g_{l_{iL}}^{(0)}(y)\,f_{e_{jR}}^{(0)}(y).\label{eoverlap}
	\end{align}
Note that the localized lepton profiles are described by real functions $g_{l_{iL}}^{(0)}{}^{\ast} = g_{l_{iL}}^{(0)}$ {and we have rotated the form of the VEV as $\langle H(y) \rangle = \left( 0,\ \langle h(y) \rangle \right)^{\mathrm{T}}$ like the SM, irrespective of $y$-dependence~\cite{Fujimoto:2013ki}.
The exact shape of the VEV is found in Eq.~(\ref{Higgs_VEV_form}).}
Here, we try to consider the configuration where the lepton doublet and the neutrino singlet profiles are extremely tightly localized around the boundaries, while the profiles of the charged lepton singlets moderately {increase} toward the $y$-positive direction.

Firstly, we mainly focus on the diagonal components of both the mass matrices.
Of particular importance of the non-diagonal terms is in discussing the structure of flavor mixing, but we postpone its detail to the next subsection.
As shown in Fig.~\ref{figure-masshierarchy_chargedlepton},
the component $m_{11}^{(e)}$ of the mass matrix for the {charged} lepton contains the overlap integral of the first generations $g_{l_{1L}}^{(0)}$ and {$f_{e_{1R}}^{(0)}$}, which live in the left side region of the extra dimension, and another one, e.g., $m_{33}^{(e)}$ contains the overlap of the third generations $g_{l_{3L}}^{(0)}$ and {$f_{e_{3R}}^{(0)}$}, which live in the right side region. Obviously, the form of the VEV {in} Eq.~(\ref{exponentialVEV}) makes a big differences {in} the overlap integrals (\ref{eoverlap}) and the exponential mass hierarchy appears for the charged {leptons}.
	\begin{align}
	m_{11}^{(e)}\ll m_{22}^{(e)}\ll m_{33}^{(e)}.
	\end{align}

{In the case of the neutrinos}, however, it is not the case.
{For the neutrinos}, we need to make the value of the bulk {masses $M_{L}$, $M_{{\cal N}}$} as large to realize an immoderate localization of the zero modes {with $M_{L}L$, $M_{{\cal N}}L$= ${\cal O}(100)$}.
In that situation, the overlap integrals in Eq.~(\ref{nuoverlap}) become {extremely} small so that tiny neutrino masses {appear} without tuning the Yukawa couplings.
Moreover, because of the immoderate localization, the effect of the $y$-dependent VEV of the gauge singlet scalar {$\langle \Phi(y) \rangle$} becomes weak, which means that non-hierarchical masses appear to the neutrinos. 
	\begin{align}
	m_{11}^{(\nu)}\lesssim m_{22}^{(\nu)}\lesssim m_{33}^{(\nu)}.
	\end{align}
{The situation is depicted in Fig.~\ref{figure-masshierarchy}.}
Thus, we can obtain {both} the charged lepton mass hierarchy and degenerated tiny neutrino masses at the same time in this model.

\subsection{Flavor mixing angles}\label{subsection-flavormixing}

In this subsection, we {discuss} the flavor mixing in our model.
Since each 5d fermion feels the point interactions at the different positions as Fig.~\ref{figure-positions}, the mass matrices {possess} the off-diagonal components. Significantly, the form of the mass matrices, which are determined by the overlap integrals {(\ref{nuoverlap})--(\ref{eoverlap})}, is restricted by the geometry of the extra dimension, {where up to three nonzero non-diagonal terms are possible.}

Generally, off-diagonal terms of a mass matrix play a very significant role in determining the corresponding mixing angles.
Also, when the magnitude of off-diagonal terms {is} sizable compared with diagonal ones, they make a primary contribution to the mass eigenvalues.
Thereby, we should carefully choose the order of the positions of the 5d fermion's point interactions.

\begin{figure}[t]
	\hspace{-5em}\includegraphics[width=1.2 \columnwidth]{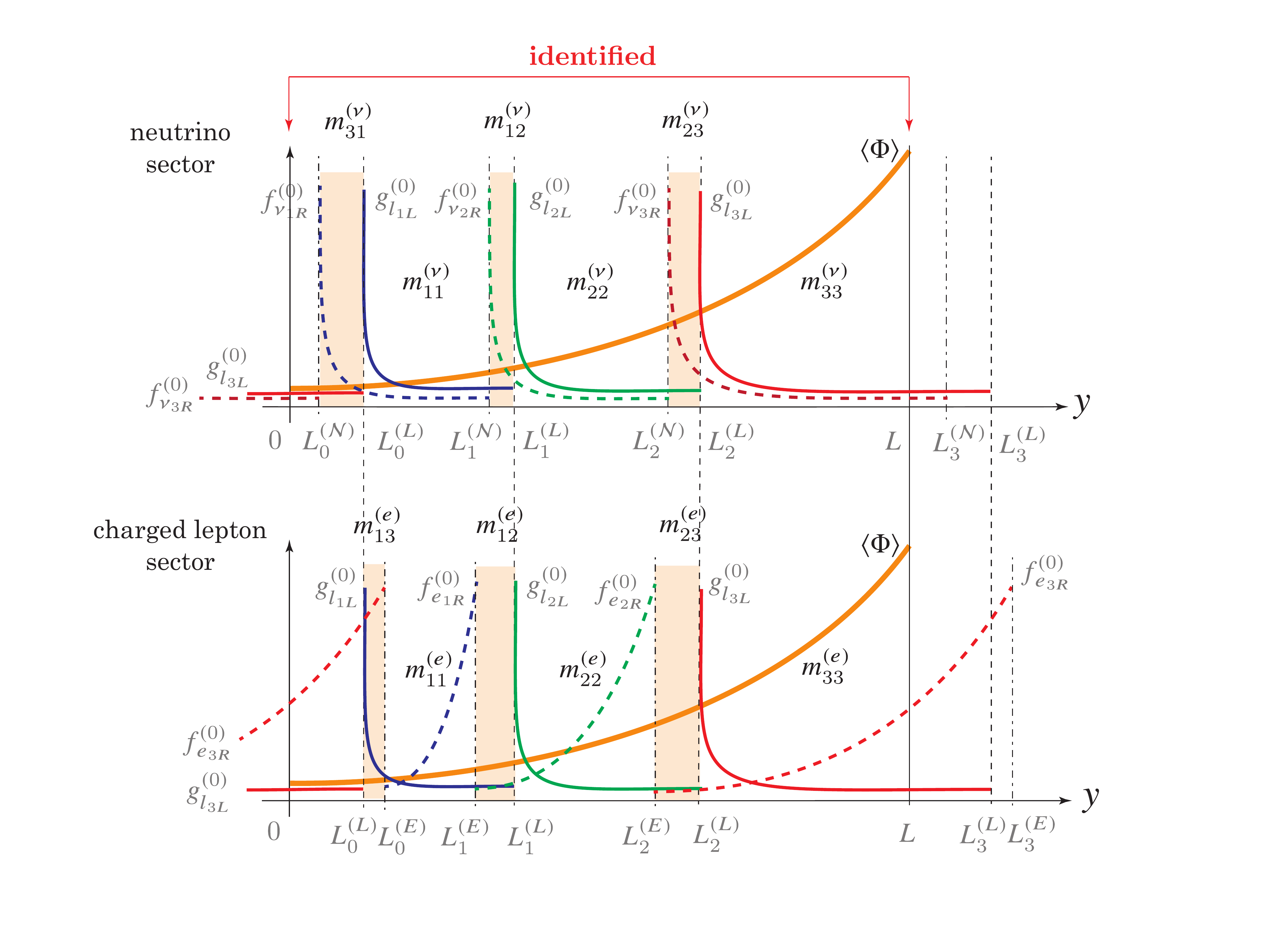}
	\vspace{-1.3cm}\caption{{\small A schematic figure of the correspondence between the components of the mass matrices and the overlap integrals. The orange colored regions indicate the off-diagonal components of the mass matrices.  $m_{ij}^{(\nu)}$ corresponds to the overlap of $g_{l_{iL}}^{(0)}$, $f_{\nu_{jR}}^{(0)}$  and $m_{ij}^{(e)}$ corresponds to {that of} $g_{l_{iL}}^{(0)}$ and $f_{e_{jR}}^{(0)}$.}}
	\label{figure-mixing}
	\end{figure}

{When we adopt the choice in Eqs.~(\ref{nuconfiguration}) and (\ref{econfiguration})}, the following three-zero textures appear:
			\begin{align}
			M^{(\nu)}=\left(\begin{array}{ccc}
			m_{11}^{(\nu)}&m_{12}^{(\nu)}&0\\[0.2cm]
			0&m_{22}^{(\nu)}&m_{23}^{(\nu)}\\[0.2cm]
			m_{31}^{(\nu)}&0&m_{33}^{(\nu)}
				\end{array}\right), 
			\hspace{3em}
			M^{(e)}=\left(\begin{array}{ccc}
			m_{11}^{(e)}&m_{12}^{(e)}&m_{13}^{(e)}\\[0.2cm]
			0&m_{22}^{(e)}&m_{23}^{(e)}\\[0.2cm]
			0&0&m_{33}^{(e)}
				\end{array}\right). \label{leptonmassmatrix}
			\end{align}
{The correspondence between the restricted mass matrix components and the configurations of the point interactions are depicted in Fig.~\ref{figure-mixing}.}
{Note that} if we change the configurations {(\ref{nuconfiguration})--(\ref{econfiguration})}, {the forms} of the mass matrices are modified.
{In general in our configuration, we cannot fill up all the off-diagonal components, but only up to three we can.}
{Therefore, the flavor mixing pattern is highly restricted. The other main possible patterns of the restricted mass matrices are represented in Appendix B.}

Now, we give some comments on the choice of the signs and magnitudes of the bulk masses {(\ref{signbulkmassL})--(\ref{signbulkmassE})} and  the configurations of the point interactions~{(\ref{nuconfiguration})--(\ref{econfiguration})}.
To produce the large mixing structure for the {leptons} with tiny neutrino masses, we need to realize the situation in which the off-diagonal components of the neutrino mass matrix are comparable with the diagonal components of that {by} the immoderate localization {via significant magnitudes of the bulk masses for ensuring the tininess of the elements}.
The choice of the sign of the bulk masses~{(\ref{signbulkmassL})--(\ref{signbulkmassN})} is really suitable for us to make the neutrino mass matrix as that situation {with the configuration of the point interactions, where all the matrix elements for the neutrinos are automatically suppressed.}
{The overlap ways of the off-diagonal components and the diagonal ones are comparable so that the large mixing of $\theta_{12}$ and $\theta_{23}$ can appear. For $\theta_{13}$, if one considers a neutrino mass matrix with $m_{13}^{(\nu)}\neq0$ (and $m_{31}^{(\nu)}=0$), the magnitude of $m_{13}^{(\nu)}$ is too small to reproduce $\theta_{13}$ due to the smallness of the overlap. This is the reason why we take the non-vanishing $m_{31}^{(\nu)}$ for the neutrino mass matrix. As a result, the choice of the configurations~{(\ref{nuconfiguration})--(\ref{econfiguration})} is favored for the realization of the leptonic mixing angles.}

{Moreover,} for the charged {leptons}, the choice of Eq.~(\ref{signbulkmassN}) leads us to the charged lepton mass hierarchy. In the configuration (\ref{signbulkmassL}) and (\ref{signbulkmassE}), the off-diagonal part of the charged lepton mass matrix is irrelevant to the lepton flavor mixing since the value of which is small due to the small overlap. In addition, the hierarchical structure of the diagonal components of the charged {leptons lessens} the effect to the flavor mixing, too. Thus, the choices {(\ref{signbulkmassL})--(\ref{signbulkmassE})} and {(\ref{nuconfiguration})--(\ref{econfiguration})} are really convenient for us to produce the flavor structure.

{Finally, we point out two issues on other possibilities.
When we only look into the charged lepton mass hierarchy, some other configurations are possible since what is required is only the primary diagonal terms and secondary off-diagonal terms in magnitude, respectively.
{On the other hand, our configuration for the neutrino sector is the most suitable one for reproducing leptonic mixing angles but there might exist an allowed region in other configurations. It would be intriguing to perform full numerical scans in all possible configurations. For the neutrino mass spectrum, our configuration predicts only the normal mass hierarchy because of the structure of $m_{11}^{(\nu)}\lesssim m_{22}^{(\nu)}\lesssim m_{33}^{(\nu)}$.}
}

\subsection{CP phase}\label{subsection-CPphase}

This subsection is devoted to the CP phase. Since, our model consists of {one-generation} fermions, the Yukawa couplings ${\cal Y}^{({\cal N})}$, ${\cal Y}^{(E)}$ cannot be a source of the physical CP phase nevertheless {they would be complex}. It was found in Ref.~\cite{Fujimoto:2013ki} that the Higgs VEV possesses the {$y$-dependent} phase under the twisted BC (\ref{HBC}), 
		\begin{align}
		\langle H(y) \rangle &=
							{\left( \begin{array}{c}
							0\\
							\langle h(y) \rangle
							\end{array}\right)}
							=
							\left( \begin{array}{c}
							0\\
							v/\sqrt{2}
							\end{array}\right)e^{i\frac{\theta}{L}y},\label{Higgs_VEV_form}\\
		{H(y)} \ &{\supset} \  {\left( \begin{array}{c}
							0\\
							v/\sqrt{2} + 1/(\sqrt{2L}) h^{(0)}
							\end{array} \right) e^{i\frac{\theta}{L}y}},\label{Higgs_physicalmode_form}
		\end{align}
{where $v\sqrt{L}$ is set as $246\,\text{GeV}$ and $\theta$ is a twist parameter.}
{$h^{(0)}$ represents the physical Higgs particle {in the unitary gauge.}}
Interestingly, the above $y$-dependent phase of the Higgs VEV can be a source of the CP phase in this case.
	\footnote{
	Here, we briefly comment on the couplings associated with ``the'' Higgs.
	At the tree level, the coupling to a SM gauge boson is completely same with that in the SM since the position dependence of $H$ is eliminated since $H$ is introduced {in} the form $(D_M H)^\dagger (D^M H)$.
	The couplings to SM fermions are also the same because the VEV and the physical {mode} profiles take the identical form as shown in Eqs.~(\ref{Higgs_VEV_form}) and (\ref{Higgs_physicalmode_form}).
	When we focus on loop-induced vertices, deviations are expected at the leading order, e.g., Higgs to diphoton coupling.
	But our model generates tree-level flavor-changing neutral current (FCNC) exchanging KK gauge bosons and
	then typically $M_{\text{KK}} \gtrsim 10^3\,\text{TeV}$ is required.
	Consequently, our Higgs boson cannot be distinguished from the SM one, unlike Universal Extra Dimension models~\cite{Petriello:2002uu,Rai:2005vy,Maru:2009cu,Nishiwaki:2011vi,Nishiwaki:2011gk,Belanger:2012mc,Kakuda:2013kba,Dey:2013cqa,Flacke:2013nta,Datta:2013xwa} inspired by the early work in the string theory context~\cite{Antoniadis:1990ew}, and is consistent with the latest LHC results.
	A challenge in phenomenology is to find an improved setup where the effect of FCNC is diminished.
	}
	
\begin{figure}[t]
	\hspace{-5em}\includegraphics[width=1.2 \columnwidth]{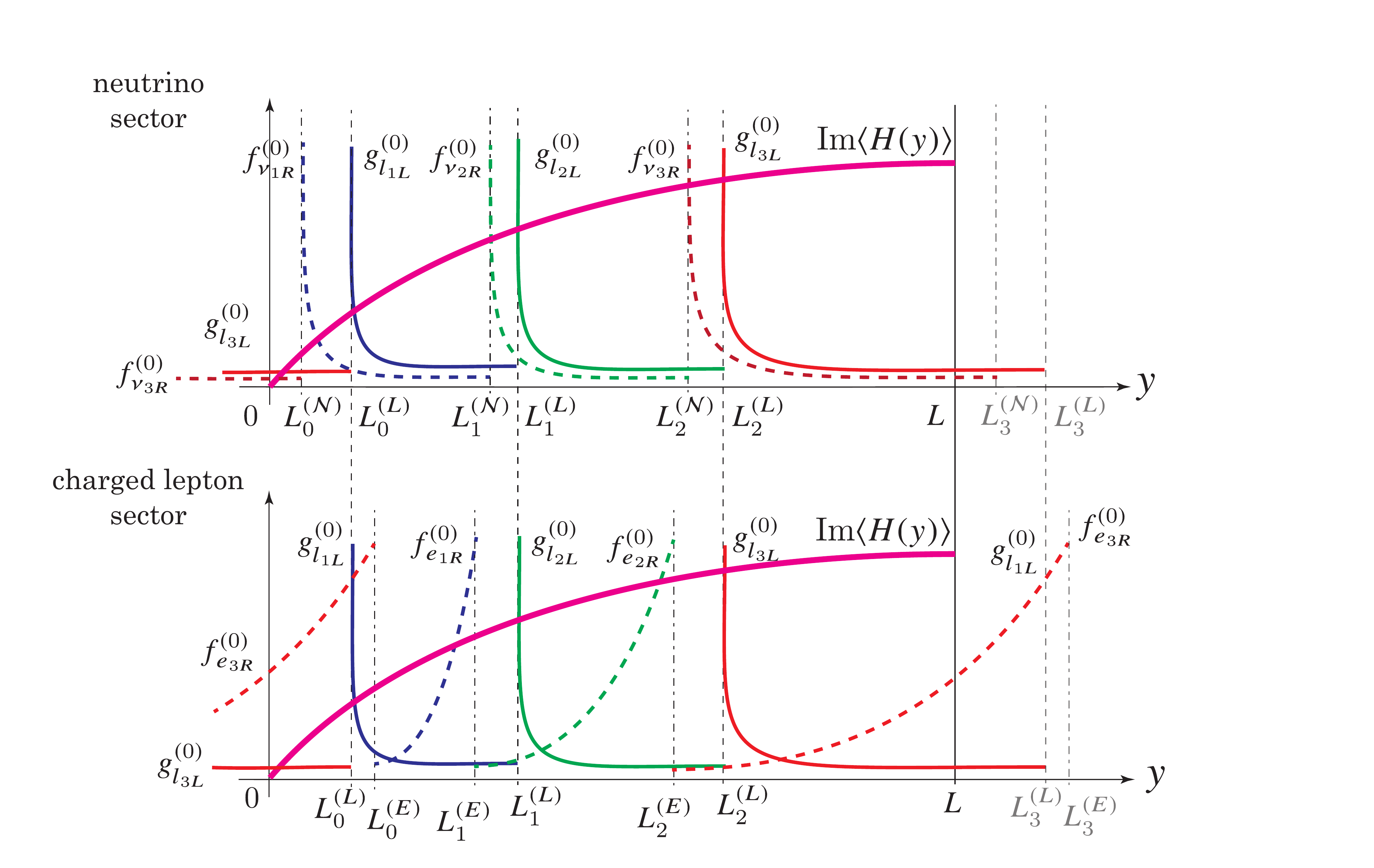}
	\vspace{-1.3cm}\caption{{\small A schematic figure of the lepton profiles and the {$y$-dependent} VEV of the Higgs doublet.}}
	\label{figure-CP}
	\end{figure}

The $y$-dependent phase of the Higgs VEV provides the different phases to the components of the mass matrix through the overlap integrals (\ref{nuoverlap})--(\ref{eoverlap}) and the physical CP phase appears {effectively}.
A schematic figure of the situation is depicted in Fig.~\ref{figure-CP}. The value of the {physical} CP phase is {a function of} the twisted parameter $\theta$ and the configuration of the extra dimension, just like the mass hierarchy. Significantly, this structure indicates that the lepton CP phase is a predictable value when we consider the quarks and leptons at the same time in this scenario {through the Yukawa couplings with the common Higgs doublet $H$}.
After fixing the value of $\theta$ to reproduce the CP phase of the CKM matrix, {we have no free parameters to tune the CP phase of the PMNS matrix.} Then, {a} full scan of the parameter space provides us a prediction for the value of the lepton CP phase. It should be investigated {in the near future}.

\subsection{Numerical analysis}\label{subsection-example}

As a typical example, we choose the parameters as
	\footnote{
	 The absolute value of the Yukawa couplings $\sqrt{|{\cal Y}^{{(\mathcal{N})}}|}$, $\sqrt{|{\cal Y}^{{(E)}}|}$, which are dimension $-1$, are $\sqrt{|{\cal Y}^{{(\mathcal{N})}}|}=0.00568 L$ and $\sqrt{|{\cal Y}^{{(E)}}|}=0.0568 L$. Obviously, there is no sizable hierarchical structure for the Yukawa couplings.}
	\begin{align}
	&\tilde{L}_{0}^{(L)}=0.2565,\hspace{2.5em}\tilde{L}_{1}^{(L)}=0.5776, \hspace{2em}\tilde{L}_{2}^{(L)}=0.9432,\nonumber\\
	&\tilde{L}_{0}^{({\cal N})}=0.08240,\hspace{1.6em} \tilde{L}_{1}^{({\cal N})}=0.3909, \hspace{1.8em}\tilde{L}_{2}^{({\cal N})}=0.7317,\nonumber\\
	&\tilde{L}_{0}^{(E)}=0.277, \hspace{2.9em} \tilde{L}_{1}^{(E)}=0.49, \hspace{3em}\tilde{L}_{2}^{(E)}=0.79,\nonumber\\[0.18cm]
	&\tilde{M}_{L}=-136.9,\hspace{1.9em} \tilde{M}_{{\cal N}}=112.1,  \hspace{1.9em}\tilde{M}_{E}=-2.00,\nonumber\\[0.2cm]
	&\tilde{M}_{\Phi}=8.67,\hspace{2.2em}  \tilde{\lambda}_{\Phi}={0.001}, \hspace{2.2em}\frac{1}{\tilde{L}_{+}}=-6.07,  \hspace{1em}\frac{1}{\tilde{L}_{-}}=8.69,\hspace{2.8em} \theta={3}, \nonumber\\[0.2cm]
	&{\tilde{{\cal Y}}^{(\mathcal{N})}}={-0.0000309-9.15\times10^{-6} \, i},\hspace{2em} {\tilde{{\cal Y}}^{(E)}}={-0.00309 - 0.000915 \,i}
	\label{lepton_parameters}
	\end{align}
where the variables with \, $\tilde{}\ $  are dimensionless parameters, which {are scaled} by using the circumference $L$ of the extra dimension.
By calculating the mass eigenvalues, the PMNS matrix and the Jarlskog parameter {in the leptonic sector} $J_{{\rm lepton}}$~\cite{Jarlskog:1985ht,Jarlskog:1985cw} through the overlap integrals {(\ref{nuoverlap})--(\ref{eoverlap})}, the following values {are} obtained.
	\begin{align}
	&{m_{\nu_1}}=0.0092 \ {\rm eV},\hspace{3.8em} {m_{\nu_2}}=0.013\  {\rm eV}, \hspace{3.4em} {m_{\nu_3}}={0.018}\ {\rm eV},\nonumber\\
	&m_{{\rm electron}}={0.519}\ {\rm MeV},\hspace{1.8em} m_{{\rm muon}}={106}\ {\rm MeV}, \hspace{2.4em}m_{{\rm tau}}={1.778}
	\ {\rm GeV},\nonumber\\[0.2cm]
	&\sin^2\theta_{12}={0.333}, \hspace{2em}\sin^2\theta_{23}={0.435}, \hspace{2em}\sin^2\theta_{13}={0.0239},\nonumber\\
	&J_{{\rm lepton}}={0.0214}\ \ (\sin\delta ={0.607}).
	\end{align}
Obviously, we obtained the neutrino mass scale $m_{\nu}\sim {\cal O}(0.01)\,{\rm eV}- {\cal O}(0.1)\,{\rm eV}$ and the charged lepton mass hierarchy at the same time. The ratio of the above results and the experimental results are shown in the follows:
	\begin{align}
	&\sqrt{\frac{\delta m^2}{\delta m^2{}^{({\rm exp.)}}}}={1.03} ,\hspace{2.8em}\sqrt{\frac{\Delta m^2}{\Delta m^2{}^{({\rm exp.)}}}}={0.285},\nonumber\\[0.3cm]
	&\frac{m_{{\rm electron}}}{m_{{\rm electron}}^{({\rm exp.})}}=1.02 ,\hspace{2.8em}\frac{m_{{\rm muon}}}{m_{{\rm muon}}^{({\rm exp.})}}=0.995, \hspace{2.6em}\frac{m_{{\rm tau}}}{m_{{\rm tau}}^{({\rm exp.})}}=1.00,\nonumber\\[0.3cm]
	&\frac{\sin^2\theta_{{\rm 12}}}{\sin^2\theta_{\rm 12}^{({\rm exp.})}}={1.08} ,\hspace{2.8em}\frac{\sin^2\theta_{{\rm 23}}}{\sin^2\theta_{\rm 23}^{({\rm exp.})}}={1.02}, \hspace{2.6em}\frac{\sin^2\theta_{{\rm 13}}}{\sin^2\theta_{\rm 13}^{({\rm exp.})}}={1.02},\label{leptonratio}
	\end{align}
where we defined $\delta m^2$ and $\Delta m^2$ as
	\begin{align}
	&\delta m^2\equiv m_{\nu_{2}}^2-m_{\nu_{1}}^2,\\[0.15cm]
	&\Delta m^2\equiv m_{\nu_{3}}^2 -\left(\frac{m_{\nu_{1}}^2-m_{\nu_{2}}^2}{2}\right),
	\end{align}
according to Ref.~\cite{Capozzi:2013csa}. The mixing angles of the PMNS matrix are within the 3$\sigma$ range \cite{Capozzi:2013csa}. 
	\footnote{
	The value of $\sqrt{\frac{\Delta m^2}{\Delta m^{2({\rm exp.})}}}$ in our example is slightly smaller than experimental range, i.e. about factor-four. For a sizable splitting in $\sqrt{\Delta m^2}$, all the components of the neutrino mass matrix should not be the same magnitude, where slight differences are required.
The $y$-dependent singlet VEV is expected to make the difference explicitly, but the amount is not enough among the second and third generations due to the extremely left-side localization of the neutrino profiles, where the effect is minimized.
Besides, when we simply try to move the positions of the point interactions to right side, the above problem can be solved, but simultaneously and inevitably, the magnitude of the $(3,1)$ component is reduced and $\theta_{13}$ gets to be too small.
An exhaustive parameter scanning would help us to find a rather reasonable point.
	}


Interestingly, we can interpret the above results instinctively from the geometry. Since we chose the sign of the bulk mass as Eqs.~{(\ref{signbulkmassL})--(\ref{signbulkmassE})}, the magnitude of the off-diagonal components can be comparable with that of the diagonal components for the neutrino mass matrix, which is an indication of the large mixing.
In addition, we put the large values on the bulk masses $M_{L},\ M_{{\cal N}}$, so that the immoderate localization of the zero modes makes the overlap integrals small. Because of this, tiny neutrino masses are realized. Moreover, large bulk masses reduce the effect of the {$y$-dependent} VEV of the gauge singlet scalar, which is a source of the mass hierarchy, via the immoderate localization.
Due to this, we obtained the degenerated neutrino masses. 
{In this way, we have chosen the geometry, i.e., the positions of the point interactions, and we have then obtained the suitable lepton flavor structure.}

Finally, we comment on the number of the {parameters}. We have fourteen parameters for the lepton flavor structure in our model, where we ignore the parameter for the exponential VEV of the gauge singlet. The {number of physical quantities is ten, where ratios to experimental values are} represented in Eq.~(\ref{leptonratio}), so that we have much more the input parameters than the outputs. Though, it does not mean that we can always reproduce the experimental results in our model. Since the restriction of the geometry is so tight as in Eq.~(\ref{leptonmassmatrix}), we cannot produce an arbitrary mixing angle nor the degenerated neutrino masses. The full scan of the parameter space is needed for finding out the full structure of our model.
This is one of the future works.

\subsection{Constraints on KK scale and cutoff scale}\label{subsection-constraint}

In this part, we would discuss the constraints on the KK scale $M_{\text{KK}}$, which is defined as $M_{\text{KK}} \simeq 1/L$, and the cutoff scale of the theory $\Lambda$.
At first, we focus on the relation between $M_{\text{KK}}$ and $\Lambda$ in a theory with a compact extra dimension.
In general above $M_{\text{KK}}$, couplings in the theory run power-like, not logarithmically in a four-dimensional theory~\cite{Dienes:1998vh,Dienes:1998vg}.
Then, a coupling rapidly blows up or blows down at an energy near $M_{\text{KK}}$, where the theory breaks down and the cutoff scale $\Lambda$ should be set here.
Typically in the 5d minimal Universal Extra Dimension when $M_{\text{KK}}$ is $\mathcal{O}(1)\,\text{TeV}$, the Higgs vacuum stability put a stringent bound on $\Lambda$ as $\Lambda \simeq 5 M_{\text{KK}}$~\cite{Bhattacharyya:2006ym,Kakuda:2013kba}.
In this analysis, we simply adapt the following relation,\footnote{
This relation is evaluated when $M_{\text{KK}}$ is $\mathcal{O}(1)\,\text{TeV}$, where we can see some dependence on the value of $M_{\text{KK}}$~\cite{Bhattacharyya:2006ym,Kakuda:2013kba}.
When $M_{\text{KK}}$ gets to be away from $\mathcal{O}(1)\,\text{TeV}$, the values of $M_{\text{KK}}$ and $\Lambda$ would become close.
In addition in our theory, a large difference in the masses of the KK particles with a same KK index would be found because of the field localization, which could alter the relation.
In this analysis, we simply ignore this issue.
}
\begin{equation}
\Lambda \simeq M_{\text{KK}}.
\label{cutoff_KKmass_relation}
\end{equation}

The constraints can be classified into two categories.
The first one is the restriction from a process which KK particles contribute to, e.g., FCNC processes, Lepton Flavor Violation (LFV) and the modification to the Newton's law of gravity.
In this case, the value of $M_{\text{KK}}$ is restricted by these phenomena.
The other possibility is the constraint originating from a gauge-invarinant higher-dimensional operator suitably suppressed by the scale $\Lambda$.
When we consider that the accidental global symmetries of the SM, e.g., baryon and lepton numbers, are not preserved above $\Lambda$, we can write down the operators causing proton decay and neutron-antineutron oscillation, and they put bounds on the possible value of $\Lambda$.
In the following part, we examine both cases.

As we mentioned, a typical bound on $M_{\text{KK}}$ from the $K$--$\overline{K}$ mixing is $M_{\text{KK}} \gtrsim \mathcal{O}(10^3)\,\text{TeV}$, which would be obtained by an operator analysis.
We also simply estimate the bounds from the LFV processes of $\mu \to e \gamma$ and $\mu \to 3e$ based on the formulas in Ref.~\cite{Agashe:2006iy} and both bounds are $M_{\text{KK}} \gtrsim \mathcal{O}(10)\,\text{TeV}$.
The constraint from the modification to the Newton's law of gravity is $R \lesssim \mathcal{O}(10^{-6})\,\text{m}$ (in the most stringent case)~\cite{Adelberger:2009zz}, {which is equivalent to $M_{\text{KK}} \gtrsim \mathcal{O}(0.1)\,\text{eV}$.
We completely neglect it in our model.}

In our 5d theory, operators generating proton decay are
\begin{equation}
QQQL,\ \mathcal{D} \mathcal{U} Q L,\ \mathcal{U} \mathcal{D} E \mathcal{U},\ QQ\mathcal{U}E,
\label{protondecay_operator}
\end{equation}
where they are dimension-eight operators since fermions hold mass dimension of two in 5d.
As mentioned before, 5d fermion fields do not have generation index, which is ``spontaneously" generated from
the KK expansion.
In our analysis, we focus on the operators including only the first-generation fermions, which give us a stringent bound. 
Interestingly in our parameter configuration describing the quarks~(\ref{quark_parameters}) and the leptons~(\ref{lepton_parameters}), there is no overlap of the first-generation singlets
{within the four operators in Eq.~(\ref{protondecay_operator})}.
Then, it is enough to evaluate the limitation from $QQQL$.

After the integration along $y$, this operator looks like a dimension-six operator as follows:
\begin{align}
\int_0^L dy \frac{c_{QQQL}}{\Lambda^3}QQQL &=
\int_0^L dy \frac{c_{QQQL}}{\Lambda^3} \epsilon_{ijk} \epsilon_{ab} \epsilon_{cd}
	\left( \overline{(Q^C)}_{ia} Q_{jc} \right)
	\left( \overline{(Q^C)}_{kb} L_{d} \right) \notag \\
&\to
\frac{c_{QQQL}}{\Lambda^3} \,I_{QQQL}\, \epsilon_{ijk} \epsilon_{ab} \epsilon_{cd}
	\left( \overline{\left(q^{(0)}_{1L}\right)^C_{ia}} \left(q^{(0)}_{1L}\right)_{jc} \right)
	\left( \overline{\left(q^{(0)}_{1L}\right)^C_{kb}} \left(l^{(0)}_{1L}\right)_{d} \right),
	\label{proton_decay_operator}
\end{align}
with the overlap integral with mass dimension of one
\begin{align}
I_{QQQL} = \int_0^{L} dy \left( g_{q^{(0)}_{1L}} \right)^3 g_{l^{(0)}_{1L}} \simeq 2.7/L .
\end{align}
Here, we pick up the part including only the first-generation fermions in the last line of Eq.~(\ref{proton_decay_operator}).
$c_{QQQL}$ is an undermined dimensionless constant, the charge conjugation $Q^C$ is defined as $Q^C := (i \Gamma^2 \Gamma^0) \left(\overline{Q}\right)^T$, $i,j,k$ and $a,b,c,d$ are the indices of the fundamental representation of the $SU(3)_C$ and $SU(2)_W$ gauge groups, respectively.
As a crude estimation, the overall coefficient part should obey the following inequality~\cite{Appelquist:2001mj},
\begin{equation}
\frac{2.7 \, c_{QQQL}}{\Lambda^3 L} \simeq \frac{2.7 \, c_{QQQL}}{\Lambda^2} \lesssim 10^{-24}\,\text{TeV}^{-2}.
\end{equation}
When we consider $c_{QQQL}$ should be natural as $\mathcal{O}(1)$, not only the cutoff scale $\Lambda$, but also the KK scale $M_{\text{KK}}$ takes the constraint through the relation in Eq.~(\ref{cutoff_KKmass_relation}) as
\begin{equation}
\Lambda,\ M_{\text{KK}} \gtrsim \mathcal{O}\left(10^{15}\right)\,\text{GeV} \simeq M_{\text{GUT}},
\end{equation}
where $M_{\text{GUT}}$ represents the typical scale of the Grand Unified Theory (GUT) since no suppression factor comes from the overlap integral in our case.
Note that we can find operators contributing to neutron-antineutron oscillation.
But the bound from this estimated based on the formula in~\cite{Mohapatra:2009wp} is $\Lambda,\, M_{\text{KK}} \gtrsim \mathcal{O}({10^2})\,\text{TeV}$, which are much less significant compared with the previous one.

In conclusion, only within the effective theory below the cutoff scale, the most stringent constraint on $M_{\text{KK}}$ is from the $K$--$\overline{K}$ mixing as $M_{\text{KK}} \gtrsim \mathcal{O}(10^3)\,\text{TeV}$.
On the other hand, including the estimation of the effects from the physics above the cutoff with the assumption that the coefficient of higher-dimensional operator is $\mathcal{O}(1)$, the result is $\Lambda,\ M_{\text{KK}} \gtrsim M_{\text{GUT}}$.\footnote{
When we assume the conservation of baryon number above the cutoff scale, proton decay never occurs and the condition, $\Lambda$ and $M_{\text{KK}} \gtrsim \mathcal{O}(10^3)\,\text{TeV}$, is enough for circumventing the experimental bounds.
}
Finally we comment on particle cosmology issues.
Due to the existence of the point interactions in the bulk space,
the translational invariance along the extra direction is
drastically violated, which means that no discrete symmetry
ensures the stability of a KK particle.
Also, no extremal suppression of the gauge couplings could be
expected in our configuration.
Thereby in our theory, no candidates of absolutely stable or
long-lived particles exists and there would be no serious
constraint from cosmogonical issues in an early stage of
the universe.\footnote{
Within a field theory with extra dimensions, no limiting temperature exists like the Hagedorn temperature in the string theory.
}

\section{{\large Summary and Discussions} \label{section:summary}}

In the part so far, we have discussed the possibility of explaining all the lepton properties in the context of the extra dimension with point interactions in the bulk space of $S^1$.
We have found a suitable parameter configuration where all the observed values are generated with acceptable precision.

The geometry of our model tells us the following tendency.
Due to the extra-dimension coordinate-dependent VEV in the gauge singlet scalar, all the three values of the neutrino masses can get to be minuscule but not zero, and the normal mass hierarchy is preferable.
The most appealing point is that, in the present configuration, the magnitude of the CP-violating effect in the lepton sector parameterized by the Jarlskog parameter is much greater than that in the quark sector. {It might be checked at T2K and NO$\nu$A experiments in the future.}
They have the common origin as the complex VEV in the Higgs doublet and thereby its value is ``predictive'' after assigning the all the parameters among the quark part.
We expect that the telltale signature could be detected in future neutrino experiments as an indirect ``proof'' of our scenario.

Now, we recognize that {our idea of point interactions is successful in describing the} flavor phenomenology of the quark and lepton sectors in the context of the $SU(2)_W \times U(1)_Y$ EW gauge theory.
One fascinating further direction is to consider a GUT model with {the} existence of point {interactions}.
In usual GUT context, the gauge structure is explained by an {extremely} sophisticated matter, but the matter configurations, especially in the number of fermions, are not so well-integrated.
Another motivation which we consider a higher-dimensional GUT {models} is that we can solve the doublet-triplet splitting problem in the Higgs sector in a natural way by twisted orbifold boundary conditions~\cite{Kawamura:1999nj,Kawamura:2000ev,Hebecker:2001jb}.

Another orientation is to estimating quantum effects with {the} existence of point {interactions}.
To understand the system more concretely,
we should evaluate possible corrections on the ``moduli'' of positions of point interactions via the Casimir force~\cite{de_Albuquerque:2003uf,Bajnok:2005dx,Pawellek:2008st} or the mass term generation for the extra-direction scalar of a 5d gauge field via the Hosotani mechanism~\cite{Hosotani:1983xw,Hosotani:1988bm,Hatanaka:1998yp}.\footnote{
{Another nontrivial topics is the structure of the boundary-localized anomaly terms~\cite{Callan:1984sa,ArkaniHamed:2001is} in a theory with multiple boundaries like our model.}
}
We will publish the work in this direction {in the near future}~\cite{Kobe:moduli}.

\section*{Acknowledgments}

K.N. would like to thank Yang Hwan Ahn, Eung Jin Chun, Rohini M. Godbole, Daisuke Harada, Hisaki Hatanaka, Dong-Won Jung, Pyungwon Ko, V. Suryanarayana Mummidi, Hiroshi Okada and Sudhir Vempati for valuable discussions.
{Especially, we would appreciate Jiro Soda and Morimitsu Tanimoto for giving us useful comments on constraints from cosmology and neutrino mass texture, respectively.}
K.N. also thank the Centre for High Energy Physics, Indian Institute of Science, Bangalore and Korea Institute for Advanced Study for warm hospitality.
K.N. is partially supported by funding available from the Department of 
Atomic Energy, Government of India for the Regional Centre for Accelerator-based
Particle Physics (RECAPP), Harish-Chandra Research Institute.
This work is supported in part by a Grant-in-Aid for Scientific Research [Grants No.\,$25\cdot3825$~(Y.F.), No.\,$22540281$ and No.\,$20540274$~(M.S.), {No.\,$24\cdot 801$~(R.T.)}] from the Japanese Ministry of Education, Science, Sports and Culture.

\appendix

\section*{Appendix}
\section{Quark flavor structure from point interactions}

\hspace{1.5em}In this appendix, we shortly review the analysis in the quark sector in Ref.~\cite{Fujimoto:2013ki}.
The basic strategy for the quark sector is the same with that for the leptons which has been explained in section~\ref{section:lepton}.
The model is given by a 5d gauge theory on a circle with point interactions. With this model, we can produce three generations of chiral massless zero modes and the quark mass hierarchy with realistic flavor mixing without breaking the higher-dimensional gauge invariance.




{In the following part, we {briefly} describe the quark part and show the results in Ref.~\cite{Fujimoto:2013ki}.}
{In} the quark sector, the point interactions provide extra boundary points to the 5d quarks and the following BC's are imposed at the points. 
	\begin{align}
	Q_{R}&=0\hspace{3em}{\rm at}\ \ \  y=L_{0}^{(Q)},L_{1}^{(Q)},L_{2}^{(Q)},L_{3}^{(Q)}\ , \label{QBC}\\
	{\cal U}_{L}&=0\hspace{3em}{\rm at}\ \ \ y=L_{0}^{({\cal U})},L_{1}^{({\cal U})},L_{2}^{({\cal U})},L_{3}^{({\cal U})}\ ,\label{UBC}\\
	{\cal D}_{L}&=0\hspace{3em}{\rm at}\ \ \ y=L_{0}^{({\cal D})},L_{1}^{({\cal D})},L_{2}^{({\cal D})},L_{3}^{({\cal D})}\ .\label{DBC}
	\end{align}
Under the BC's, three generations of chiral massless zero modes are produced from {one-generation} 5d quarks:
	\begin{align}
	Q(x,y)&=\sum_{i=1}^{3}q^{(0)}_{i L}(x)g_{q_{iL}}^{(0)}(y)+({\rm KK\ modes}),\\
	{\cal U}(x,y)&=\sum_{i=1}^{3}u^{(0)}_{i R}(x)f_{u_{iR}}^{(0)}(y)+({\rm KK\ modes}),\\
	{\cal D}(x,y)&=\sum_{i=1}^{3}d^{(0)}_{i R}(x)f_{d_{iR}}^{(0)}(y)+({\rm KK\ modes}),
	\end{align}
and the masses of the fermions are represented by the following overlap integrals.
	\begin{align}
	m_{ij}^{(u)}&={\cal Y}^{({\cal U})}\int^{L}_{0}dy \,{\langle h(y)\rangle^{\ast}} \,\langle \Phi(y)\rangle \,g_{q_{iL}}^{(0)}(y)\,f_{u_{jR}}^{(0)}(y),\label{uoverlap}\\
	m_{ij}^{(d)}&={\cal Y}^{({\cal D})}\int^{L}_{0}dy \,{\langle h(y)\rangle} \,\langle \Phi(y)\rangle \,g_{q_{iL}}^{(0)}(y)\,f_{d_{jR}}^{(0)}(y), \label{doverlap}
	\end{align}
where the VEV profiles of the Higgs doublet $H$ and the additional singlet scalar $\Phi$ are totally the same with those for the leptons as discussed in section~\ref{section:lepton}.

{In the quark part, we adopt the following ordering in the position of the point interactions for 5d quarks,}
\begin{align}
&{0\,(=L_0^{(Q)}) < L_0^{(\mathcal{U})} < L_1^{(\mathcal{U})} < L_1^{(Q)} < L_2^{(\mathcal{U})} < L_2^{(Q)} < L < L_3^{(\mathcal{U})}}, \notag \\
&{0\,(=L_0^{(Q)}) < L_0^{(\mathcal{D})} < L_1^{(\mathcal{D})} < L_1^{(Q)} < L_2^{(\mathcal{D})} < L_2^{(Q)} < L < L_3^{(\mathcal{D})}},
\end{align}
{and the signs of the quark bulk masses,}
\begin{align}
{M_Q >0,\quad M_{\mathcal{U}} <0,\quad M_{\mathcal{D}}>0},
\end{align}
{for realizing the {large/moderate} mass hierarchy in up-/down-quark sector and the small mixing angle with an CP phase, simultaneously.}
{In this setup}, the restricted mass matrices {are} given by the following {forms},
	\begin{align}
	M^{(u)}=\left(\begin{array}{ccc}
			m_{11}^{(u)}&m_{12}^{(u)}&m_{13}^{(u)}\\[0.2cm]
			0&m_{22}^{(u)}&m_{23}^{(u)}\\[0.2cm]
			0&0&m_{33}^{(u)}
				\end{array}\right), 
	\hspace{3em}
	M^{(d)}=\left(\begin{array}{ccc}
			m_{11}^{(d)}&m_{12}^{(d)}&m_{13}^{(d)}\\[0.2cm]
			0&m_{22}^{(d)}&m_{23}^{(d)}\\[0.2cm]
			0&0&m_{33}^{(d)}
				\end{array}\right). \label{quarkmassmatrix}
	\end{align}
Because of this constraint, it is highly non-trivial whether we can produce the realistic quark flavor structure in this model even though we have more input parameters than outputs.

It has been found that, at least, there is a parameter {set} in which the following physical quantities are reproduced.
	\begin{align}
	&\tilde{L}_{0}^{(Q)}=0,\hspace{2.8em}\tilde{L}_{1}^{(Q)}=0.30, \hspace{2em}\tilde{L}_{2}^{(Q)}={0.660},\nonumber\\
	&\tilde{L}_{0}^{(\mathcal{U})}=0.024,\hspace{1.1em} \tilde{L}_{1}^{(\mathcal{U})}=0.026, \hspace{1.5em}\tilde{L}_{2}^{(\mathcal{U})}=0.52,\nonumber\\
	&\tilde{L}_{0}^{(\mathcal{D})}={0.07}, \hspace{1em} \tilde{L}_{1}^{(\mathcal{D})}=0.18, \hspace{1.8em}\tilde{L}_{2}^{(\mathcal{D})}=0.646,\nonumber\\[0.18cm]
	&\tilde{M}_{Q}={6},\hspace{2.3em} \tilde{M}_{{\cal U}}={-6},  \hspace{1.7em}\tilde{M}_{{\cal D}}={5},
	\label{quark_parameters} \\
	&\tilde{M}_{\Phi}=8.67,\hspace{2.2em}  \tilde{\lambda}_{\Phi}={0.001}, \hspace{2.2em}\frac{1}{\tilde{L}_{+}}=-6.07,  \hspace{1em}\frac{1}{\tilde{L}_{-}}=8.69,\hspace{2.8em} \theta={3},    \label{HiggsSingletparameter}
	\end{align}
	\begin{align}
	&m_{{\rm up}}=2.5 \ {\rm MeV},\hspace{2.8em}m_{{\rm charm}}=1.339\  {\rm GeV}, \hspace{2em}m_{{\rm top}}=173.3\ {\rm GeV},\nonumber\\
	&m_{{\rm down}}=4.8\ {\rm MeV},\hspace{1.8em} m_{{\rm strange}}=104\ {\rm MeV}, \hspace{2.4em}m_{{\rm bottom}}=4.183\ {\rm GeV},\\[0.3cm]
	&|V_{{\rm CKM}}|=\left( \begin{array}{ccc}
				0.971&0.238&0.00377\\
				0.237& 0.971&0.0403\\
				0.00887&0.0395&0.999
					\end{array}\right),\\[0.3cm]
	&J_{{\rm quark}}=3.23\times10^{-5},
	\end{align}
where the parameters listed in Eq.~(\ref{HiggsSingletparameter}) are common values in both of the quarks and the leptons.
The ratio of the above values to the experimental values shows a good accuracy~\cite{Beringer:1900zz},
	\begin{align}
	&\frac{m_{{\rm up}}}{m_{{\rm up}}^{({\rm exp.})}}=1.07 ,\hspace{2.8em}\frac{m_{{\rm charm}}}{m_{{\rm charm}}^{({\rm exp.})}}=1.05, \hspace{2.6em}\frac{m_{{\rm top}}}{m_{{\rm top}}^{({\rm exp.})}}=1.00,\nonumber\\[0.3cm]
	&\frac{m_{{\rm down}}}{m_{{\rm down}}^{{(\rm exp.)}}}=0.993,\hspace{2.4em} \frac{m_{{\rm strange}}}{m_{{\rm strange}}^{({\rm exp.})}}=1.10, \hspace{2.4em}\frac{m_{{\rm bottom}}}{m_{{\rm bottom}}^{({\rm exp.})}}=1.00,\\[0.3cm]
	&\left|\frac{V_{{\rm CKM}}}{V_{{\rm CKM}}^{({\rm exp.)}}}\right|=\left( \begin{array}{ccc}
				0.997&1.06&1.07\\
				1.06& 0.998&0.978\\
				1.02&0.978&1.00
					\end{array}\right),\\[0.3cm]
	&\frac{J_{{\rm quark}}}{J_{{\rm quark}}^{({\rm exp.})}}=1.09.
	\end{align}
	

\section{The form of the mass matrix}
In this section, we represent the main possible mass matrix forms. $L_{i}^{(D)}$ ($i=0,1,2$) indicates the position of the point interactions for a $SU(2)$ doublet field and  $L_{i}^{(S)}$ ($i=0,1,2$) indicates the position of the point interactions for a $SU(2)$ singlet in this section.
Here, we show the criteria for classifying possible types of the mass matrix:

\begin{itemize}
\item {for $1$-$2$ mixing:
	\Bigg\{
	\begin{tabular}{ll}
	$L_{1}^{(D)} < L_{1}^{(S)}$ & : nonzero $(2,1)$, \\[2pt]
	$L_{1}^{(D)} > L_{1}^{(S)}$ & : nonzero $(1,2)$,
	\end{tabular}}
\item {for $2$-$3$ mixing:
	\Bigg\{
	\begin{tabular}{ll}
	$L_{2}^{(D)} < L_{2}^{(S)}$ & : nonzero $(3,2)$, \\[2pt]
	$L_{2}^{(D)} > L_{2}^{(S)}$ & : nonzero $(2,3)$,
	\end{tabular}}
\item {for $3$-$1$ mixing:
	\Bigg\{
	\begin{tabular}{ll}
	$L_{3}^{(D)}\,(\sim L_{0}^{(D)}) < L_{3}^{(S)}\,(\sim L_{0}^{(S)})$ & : nonzero $(1,3)$, \\[2pt]
	$L_{3}^{(D)}\,(\sim L_{0}^{(D)}) > L_{3}^{(S)}\,(\sim L_{0}^{(S)})$ & : nonzero $(3,1)$,
	\end{tabular}}
\end{itemize}
where eventually configurations are divided into eight categories. 
Note that we cannot judge whether or not the diagonal components have nonzero values only {from the above discussion}.
For example, in the configuration: $L_0^{(D)} < L_0^{(S)} < L_1^{(S)} < L_2^{(S)} < L_1^{(D)}$, the $(2,2)$ component becomes zero.
In the following list, such a possibility is not written down.

\begin{description}
\item[$\mathbf{1.}$] {$0<L_{0}^{(D)}<L_{0}^{(S)}<L_{1}^{(S)}<L_{1}^{(D)}<L_{2}^{(S)}<L_{2}^{(D)}<L$:}
	\begin{align}
	M_{({\rm I})}=\left(\begin{array}{ccc}
		m_{11}&m_{12}&m_{13}\\[0.2cm]
		0&m_{22}&m_{23}\\[0.2cm]
		0&0&m_{33}
		\end{array}\right).
	\end{align}
\item[$\mathbf{2.}$] {$0<L_{0}^{(D)}<L_{0}^{(S)}<L_{1}^{(D)}<L_{1}^{(S)}<L_{2}^{(S)}<L_{2}^{(D)}<L$:}
	\begin{align}
	M_{({\rm I\!I})}=\left(\begin{array}{ccc}
		m_{11}&0&m_{13}\\[0.2cm]
		m_{21}&m_{22}&m_{23}\\[0.2cm]
		0&0&m_{33}
		\end{array}\right).
	\end{align}
\item[$\mathbf{3.}$] {$0<L_{0}^{(D)}<L_{0}^{(S)}<L_{1}^{(S)}<L_{1}^{(D)}<L_{2}^{(D)}<L_{2}^{(S)}<L$:}	
	\begin{align}
	M_{({\rm I\!I\!I})}=\left(\begin{array}{ccc}
		m_{11}&m_{12}&m_{13}\\[0.2cm]
		0&m_{22}&0\\[0.2cm]
		0&m_{32}&m_{33}
		\end{array}\right).
	\end{align}
\item[$\mathbf{4.}$] {$0<L_{0}^{(S)}<L_{0}^{(D)}<L_{1}^{(S)}<L_{1}^{(D)}<L_{2}^{(S)}<L_{2}^{(D)}<L$:}
	\begin{align}
	M_{({\rm I\!V})}=\left(\begin{array}{ccc}
		m_{11}&m_{12}&0\\[0.2cm]
		0&m_{22}&m_{23}\\[0.2cm]
		m_{31}&0&m_{33}
		\end{array}\right).
	\end{align}
\item[$\mathbf{5.}$] {$0<L_{0}^{(D)}<L_{0}^{(S)}<L_{1}^{(D)}<L_{1}^{(S)}<L_{2}^{(D)}<L_{2}^{(S)}<L$:}
	\begin{align}
	M_{({\rm V})}=\left(\begin{array}{ccc}
		m_{11}&0&m_{13}\\[0.2cm]
		{m_{21}}&m_{22}&0\\[0.2cm]
		0&m_{32}&m_{33}
		\end{array}\right).
	\end{align}
\item[$\mathbf{6.}$] {$0<L_{0}^{(S)}<L_{0}^{(D)}<L_{1}^{(D)}<L_{1}^{(S)}<L_{2}^{(S)}<L_{2}^{(D)}<L$:}
	\begin{align}
	M_{({\rm V\!I})}=\left(\begin{array}{ccc}
		m_{11}&0&0\\[0.2cm]
		{m_{21}}&m_{22}&m_{23}\\[0.2cm]
		m_{31}&0&m_{33}
		\end{array}\right).
	\end{align}
\item[$\mathbf{7.}$] {$0<L_{0}^{(S)}<L_{0}^{(D)}<L_{1}^{(S)}<L_{1}^{(D)}<L_{2}^{(D)}<L_{2}^{(S)}<L$:}
	\begin{align}
	M_{({\rm V\! I\!I})}=\left(\begin{array}{ccc}
		m_{11}&m_{12}&0\\[0.2cm]
		0&m_{22}&0\\[0.2cm]
		m_{31}&m_{32}&m_{33}
		\end{array}\right).
	\end{align}
\item[$\mathbf{8.}$] {$0<L_{0}^{(S)}<L_{0}^{(D)}<L_{1}^{(D)}<L_{1}^{(S)}<L_{2}^{(D)}<L_{2}^{(S)}<L$:}
	\begin{align}
		M_{({\rm VI\!I\!I})}=\left(\begin{array}{ccc}
		m_{11}&0&0\\[0.2cm]
		{m_{21}}&m_{22}&0\\[0.2cm]
		m_{31}&m_{32}&m_{33}
		\end{array}\right).
	\end{align}
\end{description}

\bibliographystyle{JHEP}
\bibliography{lepton_letter.bib}
\end{document}